\documentclass[10pt,a4paper]{article}

\usepackage[cp1251]{inputenc}
\usepackage[T2A]{fontenc}
\usepackage[russian]{babel}

\usepackage{hyperref}
\usepackage[dvips]{graphicx}
\usepackage{amsmath}
\usepackage{amssymb}
\usepackage{amsxtra}
\usepackage{amsthm}
\usepackage{latexsym}
\usepackage{array}
\usepackage{multirow}
\usepackage{hhline}
\usepackage{color}

\numberwithin{equation}{section}

\newtheorem{theorem}{Теорема}

\newtheorem{remark}{Замечание}

\newcommand {\bs} {\boldsymbol}
\newcommand {\gs} {\geqslant}

\newcommand {\A} {\mathop{\rm Acc}\nolimits}
\newcommand {\rk} {\mathop{\rm rank}\nolimits}
\newcommand {\lsgn} {\mathop{\rm bsgn}\nolimits}
\newcommand {\ri} {\mathrm{i}}
\newcommand {\diag} {\mathop{\rm diag}\nolimits}
\newcommand {\mbf}[1] {\mathbf{#1}}

\newcommand {\mF}{\mathcal{F}}
\newcommand {\mB}{\mathcal{B}}
\newcommand {\mP}{\mathcal{P}}
\newcommand {\fts}[1] {{\small #1}}
\newcommand {\ts}[1] {\textsl{#1}}
\newcommand {\bT}{\mathbf{T}}
\newcommand {\bR}{\mathbb{R}}
\newcommand {\ds}{\displaystyle}
\newcommand {\mstrut}{\vphantom{\bigl(}}

\newcommand{\ru}{\rule{0pt}{12pt}}
\newcommand{\vk}{\varkappa}

\begin{document}

\noindent{УДК: 514.853; 517.938.5; 531.38}


\begin{center}

\Large{{\bf Топологический анализ и булевы функции. \\II. Приложения
к новым алгебраическим решениям}}

\vspace{5mm}

\normalsize

{\bf М.П.\,Харламов}

\vspace{4mm}

\small

Волгоградская академия государственной службы

Россия, 400131, Волгоград, ул. Гагарина, 8

E-mail: mharlamov@vags.ru
\end{center}

\begin{flushright}
{\it Получено 6 августа 2010 г.}
\end{flushright}

\vspace{3mm}

{ \footnotesize Работа является продолжением статьи автора ({\it Нелинейная динамика}, 2010, т.\,6, №\,4) и содержит приложения метода булевых функций
к исследованию допустимых областей и фазовой топологии трех
алгебраически разрешимых систем в задаче о движении волчка
Ковалевской в двойном поле сил. }

\normalsize

\vspace{3mm} Ключевые слова: алгебраическое разделение переменных,
интегральные многообразия, булевы функции, топологический анализ

\vspace{6mm}

\begin{center}

\large

{\bf M.P.~Kharlamov}

{\bf Topological analysis and Boolean functions. II. Application to
new algebraic solutions}

\normalsize
\end{center}

\vspace{3mm}

{\footnotesize This work continues the author's article in
\emph{Rus. J. Nonlinear Dynamics} (2010, v.\,6, N.\,4) and contains applications of the
Boolean functions method to investigation of the admissible regions
and the phase topology of three algebraically solvable systems in
the problem of motion of the Kowalevski top in the double force
field.

}

\vspace{3mm} Keywords: algebraic separation of variables, integral
manifolds, Boolean functions, topological analysis


\tableofcontents

\section{Введение}\label{sec1}
Статья является продолжением работы автора \cite{KhNDnew}, в которой
предложены конструктивные методы грубого топологического анализа
систем, имеющих аналитическое решение в форме алгебраических
зависимостей исходных фазовых переменных от некоторых
вспомогательных, которые в случае системы с числом степеней свободы
больше одной предполагаются разделенными. Кратко изложим необходимые
сведения и результаты из \cite{KhNDnew}.

Пусть система обыкновенных дифференциальных уравнений на
подмногообразии $\mP$ вещественного арифметического
пространства
\begin{equation}\label{neq1_1}
\dot{\mbf{x}} = {\mbf{X}}({\mbf{x}}),\quad \mbf{x} \in \mP
\end{equation}
имеет векторный первый интеграл $\mF$. Обозначая через $\mathbf{f}$
вектор произвольных постоянных, предположим, что на каждом
интегральном многообразии $\mF_{\mathbf{f}} = \{ \mathbf{x} \in \mP:
\mF(\mathbf{x})=\mathbf{f} \}$ найдены зависимости
\begin{equation}\label{neq1_2}
{\mbf{x}} = {\mbf{a}}+ \sum {\mbf{b}_j} Z_j,
\end{equation}
где ${\mbf{a}}, {\mbf{b}_j}$ --- однозначные вектор-функции от
параметров ${\mbf{f}}$ и вектора ${\mbf{s}}$ некоторых вспомогательных переменных ($\dim \{\mbf{s}\}+\dim \{\mbf{f}\}=\dim \mP$), а $Z_j$~-- произведения радикалов вида
\begin{equation}\label{neq1_3}
R_{i\gamma }  = \sqrt {\mstrut \pm (s_i  - e_\gamma)} \quad \quad
(e_\gamma \in {\mathbb{C}}),
\end{equation}
называемых базисными. Константы $e_{\gamma}$ также могут зависеть от $\mbf{f}$. Пусть вдобавок система \eqref{neq1_1} в
переменных $\mbf{s}$ принимает вид
\begin{equation}\label{neq1_4}
\phi_i(\mbf{s}) \dot s_i  = \sqrt {\mstrut V_i (s_i; \mbf{f})}\, , \qquad i=1,\ldots,\dim\{\mbf{s}\},
\end{equation}
где все $V_i$ --- многочлены от одной переменной $s$, имеющие корни
из совокупности $\{e_\gamma\}$, $\phi_i(\mbf{s})$ --- однозначные функции, не обращающиеся в нуль для почти всех $\mathbf{f}$. В этом случае будем говорить, что
система \eqref{neq1_1} алгебраически разрешима, а уравнения
\eqref{neq1_2}, \eqref{neq1_4} представляют собой ее алгебраическое
решение. Многочлен $V$, имеющий своими корнями все числа из совокупности
$\{e_\gamma\}$, называем максимальным. Обычно $V$ --- наименьшее общее кратное многочленов $V_i$.

Для каждого допустимого $\mathbf{f}$ (набора констант первых
интегралов, для которого $\mF_{\mathbf{f}} \ne \varnothing $)
достижимой областью $\A(\mathbf{f})$ называется множество в
пространстве переменных $\mbf{s}$, на котором выражения
\eqref{neq1_2} приводят к вещественным значениям. Любая связная
компонента $\Pi$ множества $\A(\mathbf{f})$ имеет форму
прямого произведения сегментов (отрезков, замкнутых полупрямых или
прямых). Предположим пока, что все величины $e_\gamma$
вещественны\footnote{Если $e_\gamma$ и $e_\delta$ комплексно
сопряжены, то просто заменим каждую пару соответствующих радикалов
на их произведение $R_{i\gamma}R_{i\delta}$. Подробно этот случай
разбирается в \cite{KhNDnew}. Примеры рассмотрим ниже.}, фиксируем
$\Pi$ и выберем знаки под корнем в \eqref{neq1_3} так, чтобы все
алгебраические радикалы $R_{i\gamma}$ также были вещественны.
Объединим все $R_{i\gamma}$ в один вектор $(U_1,\ldots, U_N)$ и
запишем
\begin{equation}\label{neq1_5}
Z_j=\prod_{i=1}^N U_i^{c_{ij}}, \qquad c_{ij}\in \mB=\{0,1\}.
\end{equation}

Булевым знаком назовем функцию $\lsgn : \bR \backslash \{0\} \to \mB$, такую, что
\begin{equation}\notag
\lsgn (\theta ) = \left\{
\begin{array}{l}
{0,\quad \theta  >  0}  \\ {1,\quad \theta  < 0}
\end{array}
\right. , \qquad \lsgn (\theta _1 \theta _2 ) = \lsgn (\theta _1 )
\oplus \lsgn (\theta _2 )
\end{equation}
($\oplus$ --- сумма по модулю 2). Значение в нуле несущественно.

Определим булевы переменные
\begin{equation}\label{neq1_6}
z_j=\lsgn Z_j, \qquad u_i = \lsgn U_i,
\end{equation}
тогда соотношения \eqref{neq1_5} примут вид линейной булевой
вектор-функции
\begin{equation}\label{neq1_7}
\mbf{z}=C \mbf{u}, \qquad \mbf{u} \in \mB^m, \qquad \mbf{z} \in
\mB^k
\end{equation}
с двоичной матрицей $C$ (множество $\mB$ рассматривается как поле
$\mathbb{Z}_2$).

Для фиксированного множества $\Pi$ булевы аргументы $u_i$
разбиваются на две группы. К первой группе относим булевы знаки
радикалов, не обращающихся в ноль на траекториях, лежащих в $\Pi$.
Эти аргументы объединяем в вектор ${\mbf{v}}\in \mB^m$. Во вторую
группу включим булевы знаки радикалов, периодически обращающихся в
ноль на траекториях из $\Pi$. Вектор таких аргументов обозначим
через ${\mbf{w}}\in \mB^n$. Возможность такого разбиения следует из
вида уравнений \eqref{neq1_4}. Тогда функция \eqref{neq1_7} есть
$\mB$-линейное отображение
\begin{equation}\label{neq1_8}
{C}: \mB^m \times \mB^n \to \mB^k.
\end{equation}
Введем на $\mB^m$ отношение эквивалентности относительно $C$,
полагая элементы $\mbf{v}',\mbf{v}'' \in \mB^m$ эквивалентными, если
существуют $\mbf{w}',\mbf{w}'' \in \mB^n$ такие, что
${C}({\mbf{v}'},{\mbf{w}'}) = {C}({\mbf{v}''},{\mbf{w}''})$.

\begin{theorem}\label{theo1}\emph{\cite{KhNDnew}}
Количество связных компонент многообразия
$\mathcal{F}_{\mbf{f}} $, накрывающих множество $\Pi$, равно
количеству классов эквивалентности в множестве $\mB^m$ по отношению
эквивалентности относительно функции ${C}$.
\end{theorem}

Линейные преобразования матрицы $C$, которые могут, в принципе,
изменять размерности сомножителей в пространстве-прообразе (не
<<перемешивая>> этих сомножителей) и размерность
пространства-образа, называем эквивалентными, если они не изменяют
количество классов эквивалентности в первом сомножителе.

\begin{theorem}\label{theo2}\emph{\cite{KhNDnew}}
Следующие преобразования являются эквивалентными:

$(i)$ перестановка строк;

$(ii)$ перестановка столбцов в пределах одной группы;

$(iii)$ прибавление $($по модулю $2)$ к некоторой строке другой
строки;

$(iv)$ прибавление $($по модулю $2)$ к некоторому столбцу другого
столбца той же группы;

$(v)$ отбрасывание нулевого столбца;

$(vi)$ отбрасывание нулевой строки;

$(vii)$ отбрасывание набора строк и набора столбцов второй группы, в
которых отличные от нуля элементы лежат только в их пересечении;

$(viii)$ отбрасывание строки с номером $i$ и столбца второй группы с
номером $j$, если этот столбец единичный с единицей в строке $i$.
\end{theorem}

Доказательство представлено в \cite{KhNDnew} в виде
последовательности лемм, удобных для практического использования. В
целом оно отражает следующий факт.
\begin{theorem}\label{theo3}
Выбирая подходящие базисы в $\mB^m,\mB^n,\mB^k$ и отбрасывая нулевые
строки и столбцы, матрицу $C$ можно привести к виду:
\begin{equation}\notag
C=
\begin{tabular}{||m{0.5cm}|c|m{0.5cm}||}
\multirow{2}{0.5cm}{\centering $E_P$}& \centering $0$ &
\multirow{2}{0.5cm}{
\centering $0$} \\
\hhline{||~|-|~||}
& \centering $E_Q$ & {} \\
\hhline{||-|-|-||} \centering $0$ & \centering $0$ & \centering
$E_R$
\end{tabular}\, ,
\end{equation}
где первые $P$ столбцов относятся к первой группе, остальные --- ко
второй. Количество классов эквивалентности в $\mB^m$ относительно
${C}$ равно $2^{P-Q}$.
\end{theorem}
При таком приведении используются лишь преобразования $(i)$--$(vi)$.
После этого уже очевидно, что множество классов эквивалентности в
$\mB^P$ изоморфно фактор-пространству $\mB^P/\mB^Q$. Действительно,
преобразование $(vii)$ позволяет отбросить строки и столбцы с блоком
$E_R$, а последовательность преобразований $(viii)$ удаляет блок
$E_Q$. Оставшаяся матрица $E_{P-Q}$ есть изоморфизм пространства
аргументов только первой группы --- все классы эквивалентности
состоят из одного элемента, а их количество равно $2^{P-Q}$. Это и
есть количество связных компонент интегрального многообразия
$\mF_{\mathbf{f}}$, накрывающих выбранный <<прямоугольник>> $\Pi$ в
составе $\A(\mathbf{f})$.

Решение задачи можно существенно упростить, если удается указать
некоторые мономы от базисных радикалов, через которые полностью
выражаются все $Z_j$. Тогда сами эти мономы можно принять за
базисные радикалы, что сразу же понижает размерность пространств.

\begin{theorem}\label{theo4}\emph{\cite{KhNDnew}}
Пусть ${{X}: \mB^m  \to \mB^{m_1}}$, ${{Y}:\mB^m\times \mB^n \to
\mB^{n_1}}$, ${{B}: \mB^{m_1}\times \mB^{n_1} \to \mB^k}$ -- линейные булевы век\-тор-функции, а функция ${C}$ представима в виде
${C}={B}\circ ({X},{Y})$. Предположим, что отображение ${X}$
сюръективно и для любого $\mbf{v}\in \mB^m$ сюръективно
индуцированное отображение $ {Y}(\mbf{v}, {\bs \cdot}) : \mB^n \to
\mB^{n_1}$. Тогда количество классов эквивалентности в $\mB^m$
относительно ${C}$ равно количеству классов эквивалентности в
$\mB^{m_1}$ относительно ${B}$.
\end{theorem}

Напомним также, что отображение \eqref{neq1_8} может быть применено
и к нахождению самих допустимых областей. Если зависимости
\eqref{neq1_2} записаны так, что все коэффициенты вещественны, то
условие вещественности $\mbf{x}$ имеет вид $Z_j^2 \gs 0 \, \forall
j$. Поэтому, полагая вместо \eqref{neq1_6}
\begin{equation}\notag
z_j=\lsgn Z_j^2, \qquad u_i = \lsgn U_i^2,
\end{equation}
будем иметь для той же матрицы $C$, что и в \eqref{neq1_7},
\begin{equation}\label{neq1_9}
C \mbf{u} = \mbf{z}_0,
\end{equation}
где в этом случае $\mbf{z}_0$ --- булев вектор, состоящий из нулей.
Если же в общем виде приводить все коэффициенты к вещественной форме
неудобно, то получится другое, но вполне определенное значение
$\mbf{z}_0$. Решения этой линейной системы дают явные ограничения на знаки подкоренных выражений в радикалах $U_i$. Из этих ограничений и определяется область $\A(\mathbf{f})$. Если вдобавок имеются известные условия на
расположение корней $e_\gamma$ или на выбор переменных разделения,
гарантирующие неравенство $U^2_{i_1} \gs U^2_{i_2}$, то, очевидно,
$(u_{i_1} \to u_{i_2})=1$ (стрелкой обозначена импликация). Добавив
к отображению $C$ компоненты-импликации, получим расширенное
отображение $\tilde C: \mB^m{\times}\mB^n \to \mB^{k'}$ с $k'>k$.
Условие ${\tilde C \mbf{u} = (\mbf{z}_0, \mbf{1})}$ сужает
количество систем неравенств, подлежащих решению, делая это решение
зачастую очевидным.

Продемонстрируем приведенную выше технику на трех допускающих
алгебраические решения подсистемах в задаче о движении волчка
Ковалевской в двойном поле, полная интегрируемость которой доказана
в \cite{ReySem, ReySemRus}, но явные решения получены лишь на некоторых
подмногообразиях в шестимерном фазовом пространстве.

\section{Уравнения движения и критические подсистемы}\label{sec2}
Уравнения движения волчка
Ковалевской--Реймана--Семенова-Тян-Шанского приводятся к виду
\begin{equation}\notag
\begin{array}{l}
\mbf{I}\dot{\bs \omega}=\mbf{I}{\bs \omega}\times {\bs
\omega}+\mbf{e}_1 \times {\bs \alpha}+\mbf{e}_2 \times {\bs
\beta},\qquad  \dot {\bs \alpha}= {\bs \alpha}\times{\bs \omega},
\qquad \dot{\bs \beta}= {\bs \beta}\times {\bs \omega},
\end{array}
\end{equation}
где $\mbf{I}=\diag \{2,2,1\}$, $\mbf{e}_1=(1,0,0)$,
$\mbf{e}_2=(0,1,0)$, напряженности ${\bs \alpha}$, ${\bs \beta}$
силовых полей постоянны в инерциальном пространстве и взаимно
ортогональны
\begin{equation}\notag
{\bs \alpha}^2=a^2,\qquad  {\bs \beta}^2=b^2, \qquad {\bs
\alpha}\cdot {\bs \beta}=0\qquad (a>b>0).
\end{equation}
На шестимерном фазовом пространстве $\mP \subset \bR^9({\bs
\alpha},{\bs \beta},{\bs \omega})$ определено интегральное
отображение
\begin{equation}\notag
H \times K \times G: \mP \to \bR^3,
\end{equation}
где
\begin{equation}\notag
\begin{array}{l}
H = \omega _1^2  + \omega _2^2  + \frac{1}{2}\omega _3^2 - (\alpha
_1  + \beta _2 ), \\[2mm]
K = (\omega _1^2  - \omega _2^2  + \alpha _1 - \beta _2 )^2  +
(2\omega _1 \omega _2  + \alpha _2  + \beta _1 )^2 , \\[2mm]
G = (\alpha _1 \omega _1  + \alpha _2 \omega _2  + \frac{1}{2}\alpha
_3 \omega _3 )^2  + (\beta _1 \omega _1  + \beta _2 \omega _2  + \frac{1}{2}\beta _3 \omega _3 )^2  +  \\
\qquad {} + \omega _3 (\gamma _1 \omega _1  + \gamma _2 \omega _2 +
\frac{1}{2}\gamma _3 \omega _3 ) - \alpha _1 b^2  - \beta _2 a^2.
\end{array}
\end{equation}
Здесь $\gamma _i$ --- компоненты вектора $ {\bs{\alpha }} \times
{\bs{\beta }}$.

В общем случае явное интегрирование полученной системы с тремя
степенями свободы не выполнено. Алгебраические решения получены в
следующих трех случаях:

1) подсистема с одной степенью свободы, состоящая из трех семейств
критических траекторий в первой критической подсистеме с двумя
степенями свободы --- случае Богоявленского (сама система открыта в
\cite{BogRus1,BogRus2}, существование семейств доказано в
\cite{ZotRCD,ZotRus}, где получены и уравнения формируемого ими многообразия, интегрирование выполнено в \cite{Kh361});

2) вторая критическая подсистема с двумя степенями свободы (найдена в
\cite{Odin}, разделение переменных указано в \cite{KhSavDan});

3) третья критическая подсистема с двумя степенями свободы (найдена
в \cite{Kh34}, разделение переменных указано в \cite{KhND06},
алгебраическое решение построено в \cite{KhRCD09}).

Приведем уравнения, необходимые для дальнейшего.

Замена переменных ($\ri^2=-1$)
\begin{equation}\label{neq2_1}
\begin{array}{l}
x_1 = (\alpha_1  - \beta_2) + \ri (\alpha_2  + \beta_1),\quad
x_2 = (\alpha_1  - \beta_2) - \ri (\alpha_2  + \beta_1 ), \\
y_1 = (\alpha_1  + \beta_2) + \ri (\alpha_2  - \beta_1), \quad y_2 =
(\alpha_1  + \beta_2) -
\ri (\alpha_2  - \beta_1), \\
 z_1 = \alpha_3  + \ri \beta_3, \quad
z_2 = \alpha_3  - \ri \beta_3,\\
w_1 = \omega_1  + \ri \omega_2 , \quad w_2 = \omega_1  - \ri
\omega_2, \quad w_3 = \omega_3
\end{array}
\end{equation}
позволяет компактно описать точки зависимости интегралов.
Ограничения на ${\bs \alpha}, {\bs \beta}$ примут вид
\begin{equation}\label{neq2_2}
\begin{array}{c}
z_1^2  + x_1 y_2  = r^2 ,\quad z_2^2  + x_2 y_1  = r^2 , \quad  x_1
x_2 + y_1 y_2  + 2z_1 z_2  = 2p^2 .
\end{array}
\end{equation}
Здесь и далее $p^2  = a^2  + b^2$, $r^2  = a^2  - b^2$ ($p>r>0$).

Первая критическая подсистема $\mathfrak{M}\subset \mP$
задана соотношениями
\begin{equation}\label{neq2_3}
w_1^2+x_1  = 0,\qquad w_2^2+x_2  = 0.
\end{equation}
Независимыми интегралами на $\mathfrak{M}$ выбирают $H$ и интеграл
Богоявленского
\begin{equation}\label{neq2_4}
F = w_1 w_2 w_3+z_2 w_1+z_1 w_2.
\end{equation}
Вторая критическая подсистема $\mathfrak{N} \subset \mP$
определена уравнениями
\begin{equation}\label{neq2_5}
\sqrt{x_1 x_2} w_3  - \displaystyle{\frac{x_2 z_1 w_1  + x_1 z_2 w_2}{\sqrt{x_1 x_2}}}  = 0, \qquad
\displaystyle{\frac{x_2}{x_1}Z_1-\frac{x_1}{x_2}Z_2}  = 0.
\end{equation}
В качестве независимых интегралов можно взять
\begin{equation}\label{neq2_6}
\displaystyle{M = \frac{1} {2
r^2}(\frac{x_2}{x_1}Z_1+\frac{x_1}{x_2}Z_2)}, \quad L = \frac{1}
{{\sqrt {x_1 x_2 } }}[w_1 w_2  + {{x_1 x_2  + z_1 z_2 }} M].
\end{equation}
Третья критическая подсистема $\mathfrak{O} \subset \mP$
описывается уравнениями
\begin{equation}\label{neq2_7}
\begin{array}{l}
\displaystyle{\frac{w_2 x_1+w_1 y_2+w_3 z_1}{w_1}-\frac{w_1 x_2+w_2
y_1+w_3
z_2}{w_2}=0,} \\
\displaystyle{(w_2 z_1+w_1 z_2)w_3^2+\Bigl[\frac{w_2
z_1^2}{w_1}+\frac{w_1 z_2^2}{w_2}+w_1 w_2(y_1+y_2)+ x_1 w_2^2+x_2 w_1^2\Bigr]w_3 +}\\
\qquad  + \displaystyle{ \frac{w_2^2 x_1 z_1}{w_1} + \frac{w_1^2 x_2
z_2}{w_2}+x_1 z_2 w_2+ x_2 z_1 w_1 +(w_1 z_2-w_2 z_1)(y_1-y_2)=0.}
\end{array}
\end{equation}
Здесь в качестве независимых интегралов выступают
\begin{equation}\label{neq2_8}
\begin{array}{l}
\displaystyle{S=-\frac{1}{4} \big( \frac {y_2 w_1+x_1 w_2+z_1
w_3}{w_1}+\frac{x_2 w_1+y_1 w_2+z_2 w_3}{w_2} \big),}
\\[3mm]
\displaystyle{{\rm T}=\frac{1}{2}[w_1(x_2 w_1+y_1 w_2+z_2
w_3)+w_2(y_2 w_1+x_1 w_2+z_1 w_3)]+x_1 x_2+z_1 z_2}.
\end{array}
\end{equation}
Соотношения \eqref{neq2_3}, \eqref{neq2_5}, \eqref{neq2_7} записаны в такой форме, чтобы производная по времени каждой функции в левой части была пропорциональна другой функции в этой паре.

\section{Особые периодические решения\mbox{\ } \\ и топология первой критической подсистемы}\label{sec3}

Как отмечалось, топологический анализ первой критической подсистемы
выполнен в работе~\cite{ZotRCD}, положившей начало качественным исследованиям волчка в двойном поле. Более подробно
результаты этой работы изложены в \cite{ZotRus}. Покажем, как
применить здесь алгебраические решения и булевы функции.

Бифуркационная диаграмма интегралов $H$ и $F$ есть множество решений
уравнения \cite{ZotRus}
\begin{equation}\notag
\left\{27 f^4 + 4 z [9 (f^2 h - r^4) + 2 z^2]\right\}^2-64 \left[3 (f^2 h - r^4) + z^2\right]^3 =0,
\end{equation}
где $z=h^2-2p^2$. Известна параметризация решений этого уравнения в виде кривых \cite{Kh361}
\begin{equation}\label{neq3_1}
\begin{array}{l}
\delta_1: \left\{
\begin{array}{l}
\displaystyle{h=2s-\frac{1}{s}\sqrt{(a^2-s^2)(b^2-s^2)}   }\\[2mm]
\displaystyle{f=
\sqrt{-\frac{2}{s}\sqrt{(a^2-s^2)(b^2-s^2)}}(\sqrt{a^2-s^2}+
\sqrt{b^2-s^2})}\\[3mm]
\displaystyle{\tau=(\sqrt{a^2-s^2}+\sqrt{b^2-s^2})^2}
\end{array} \right. , \quad s \in [-b,0); \\
\delta_2: \left\{
\begin{array}{l}
\displaystyle{h=2s+\frac{1}{s}\sqrt{(a^2-s^2)(b^2-s^2)}   }\\[2mm]
\displaystyle{f=\sqrt{\frac{2}{s}\sqrt{(a^2-s^2)(b^2-s^2)}}(\sqrt{a^2-s^2}-
\sqrt{b^2-s^2})}\\[3mm]
\displaystyle{\tau=(\sqrt{a^2-s^2}-\sqrt{b^2-s^2})^2}
\end{array} \right. , \quad s \in (0,b]; \\
\delta_3: \left\{
\begin{array}{l}
\displaystyle{h=2s-\frac{1}{s}\sqrt{(s^2-a^2)(s^2-b^2)}   }\\[2mm]
\displaystyle{f=\sqrt{\frac{2}{s}\sqrt{(s^2-a^2)(s^2-b^2)}}(\sqrt{s^2-b^2}-
\sqrt{s^2-a^2})}\\[3mm]
\displaystyle{\tau=-(\sqrt{s^2-b^2}-\sqrt{s^2-a^2})^2}
\end{array} \right. , \quad s \in [a,+\infty).
\end{array}
\end{equation}
Константы $s,\tau$ отвечают значениям функций (\ref{neq2_8}) на
пересечении $\mathfrak{M} \cap \mathfrak{O}$. Считаем $s$
независимым интегральным параметром. Точки функциональной
зависимости интегралов $H$ и $F$ на $\mathfrak{M}$, образующие
подсистему с одной степенью свободы, соответственно \eqref{neq3_1}
организованы в три подсистемы, которые также обозначим через $\delta_i$. Их алгебраическое решение построено в
\cite{Kh361} и имеет общий вид
\begin{equation}\label{neq3_2}
\begin{array}{ll}
\displaystyle{x_1 = \frac{(r_1+r_2)^2}{2 r^2 s} \, [\sqrt{r_2
\varphi_1(\xi)}-\sqrt{r_1 \varphi_2(\xi)}\,]^2,} & \displaystyle{x_2
= \frac{(r_1+r_2)^2}{2 r^2 s} \, [\sqrt{r_2
\varphi_1(\xi)}+\sqrt{r_1 \varphi_2(\xi)}\,]^2,}\\[3mm]
\displaystyle{y_1 = \frac{1}{2s} \, [\sqrt{r_2
\varphi_1(\xi)}-\sqrt{r_1 \varphi_2(\xi)}\,]^2 -2s,} &
\displaystyle{y_2 = \frac{1}{2s} \, [\sqrt{r_2
\varphi_1(\xi)}+\sqrt{r_1 \varphi_2(\xi)}\,]^2 -2s,}\\[3mm]
\displaystyle{z_1 = \frac{r_1+r_2}{r} \, [\sqrt{r_1
\varphi_1(\xi)}-\sqrt{r_2 \varphi_2(\xi)}\,],} & \displaystyle{z_2 =
\frac{r_1+r_2}{r}  \, [\sqrt{r_1
\varphi_1(\xi)}+\sqrt{r_2 \varphi_2(\xi)}\,],}\\[3mm]
\displaystyle{w_1 = \frac{(r_1+r_2)i}{r\sqrt{2s}} \, [\sqrt{r_2
\varphi_1(\xi)}-\sqrt{r_1 \varphi_2(\xi)}\,],} & \displaystyle{w_2 =
\frac{(r_1+r_2)i}{r\sqrt{2s}}  \, [\sqrt{r_2
\varphi_1(\xi)}+\sqrt{r_1 \varphi_2(\xi)}\,],}\\[3mm]
\displaystyle{w_3=\sqrt{-\frac{2r_1 r_2}{s}} \, \xi,} &{}
\end{array}
\end{equation}
где
\begin{equation}\notag
\begin{array}{c}
r_1^2  = a^2  - s^2 ,\qquad r_2^2  = b^2  - s^2 , \qquad
f=\ds{\sqrt{-\frac{2 r_1 r_2}{s}}(r_1+r_2)},
   \qquad \xi = \displaystyle{\frac{w_1 w_2 }
 {r_1  + r_2 }},\\
\varphi _i (\xi ) = r_i (1 - \xi ^2 ) - 2s\xi \qquad (i = 1,2),
\end{array}
\end{equation}
а $\xi (t)$ определяется уравнением
\begin{equation}\notag
\dot{\xi}  = \ds{\sqrt {\frac{1}{2s} \varphi _1 (\xi )\varphi _2
(\xi )}}.
\end{equation}
Чтобы удовлетворить (\ref{neq3_1}), необходимо выбрать параметры $r_1
,r_2$ следующим образом (считаем для определенности $f > 0$ в силу
очевидной симметрии):
\begin{equation}\notag
\begin{array}{lll}
s \in [ - b,0) & \Rightarrow & r_1  = \sqrt {a^2  - s^2}  > 0, \quad
r_2 = \sqrt {b^2  - s^2}\geqslant 0,\\
s \in (0,b]& \Rightarrow & r_1  = \sqrt {a^2  - s^2 }  > 0, \quad
r_2  =  - \sqrt {b^2  -
s^2 }  \leqslant 0,\\
s \in [a, + \infty ) & \Rightarrow & r_1  = \ri \, r_1^*, \quad r_2
= - \ri \, r_2^*; \quad 0 \leqslant r_1^*  = \sqrt {s^2  - a^2 }  <
r_2^*  = \sqrt {s^2  - b^2 } .
\end{array}
\end{equation}

В представленных алгебраических выражениях переменная $\xi $ может
быть как вещественной, так и чисто мнимой. Чтобы этого избежать,
возьмем в качестве вспомогательной переменной
\begin{equation}\notag
\eta  = \omega _1^2  + \omega _2^2  = w_1 w_2  = (r_1  + r_2 )\xi
\geqslant 0.
\end{equation}
Она фигурирует и в работе \cite{ZotRCD}, но последующие зависимости
выписаны там в тригонометрических функциях. Из (\ref{neq3_2}),
(\ref{neq2_1}) и (\ref{neq2_5}) найдем следующие алгебраические
выражения, определяющие исходные вещественные фазовые
переменные:
\begin{equation}\label{neq3_3}
\begin{array}{lll}
\ds{\alpha_1=-s-\frac{r_1^2r_2(\psi_1+\psi_2)}{2r^2s(r_1+r_2)},} &
\ds{\alpha_2=-\ri \frac{r_1^2 r_2 \sqrt{ \psi_1 \psi_2}}{r^2
s(r_1+r_2)},} & \ds{\alpha_3=\ri \frac{r_1 \sqrt{
\psi_1}}{r},}\\
\ds{\beta_1=-\ri \frac{r_1 r_2^2 \sqrt{ \psi_1 \psi_2}}{r^2
s(r_1+r_2)},} &\ds{\beta_2=-s+\frac{r_1
r_2^2(\psi_1+\psi_2)}{2r^2s(r_1+r_2)},} &
 \ds{\beta_3=- \frac{r_2\sqrt{
\psi_2}}{r},}\\
\ds{\omega_1=\ri \frac{f \sqrt{ \psi_1}}{(r_1+r_2)r},} &
\ds{\omega_2=- \frac{f \sqrt{ \psi_2}}{(r_1+r_2)r},} & \ds{\omega_3=
\frac{2f \eta}{(r_1+r_2)^2}.}
\end{array}
\end{equation}
Здесь
\begin{equation}\notag
\psi _j (\eta ) = \eta ^2  - 2s \frac{(r_1  + r_2 )\eta }{r_j } -
(r_1  + r_2 )^2 \quad (j = 1,2).
\end{equation}
Тогда
\begin{equation}\notag
\dot \eta  = \ds{\frac{1}{r_1  + r_2} \sqrt{\frac {r_1 r_2 } {2 s}
\psi_1 (\eta )\psi _2 (\eta )}}.
\end{equation}

Вводя обозначения для корней максимального многочлена $V(\eta ) =
\psi _1 (\eta )\psi _2 (\eta )$
\begin{equation}\label{neq3_4}
\begin{array}{ll}
\psi _1  = (\eta  - e'_ +  )(\eta  - e'_ -  ), & e'_ \pm   =
\displaystyle{\frac{r_1  + r_2 } {r_1 }(s \pm a)},
\\[3mm]
\psi _2  = (\eta  - e''_ +  )(\eta  - e''_ -  ), & e''_ \pm   =
\displaystyle{\frac{r_1  + r_2 } {r_1 }(s \pm b),}
\end{array}
\end{equation}
определим базисные радикалы
$$
\begin{array}{llll}
R_1  = \sqrt {\eta  - e'_ +  }\, , & R_2  = \sqrt {\eta  - e'_ -
}\,, & R_3  = \sqrt {\eta  - e''_ +  }\,, & R_4  = \sqrt {\eta  -
e''_ - }\,
\end{array}
$$
и булевы аргументы
$$
u_\gamma   = \lsgn R_\gamma \quad (\gamma  = 1,\ldots, 4).
$$
Булева вектор-функция $ C: \mB^4  \to \mB^2$, описывающая
надстройку, имеет компоненты
\begin{equation}\label{neq3_5}
z_1  = u_1  \oplus u_2, \quad z_2  = u_3  \oplus u_4,
\end{equation}
так как отображение (\ref{neq3_3}) содержит только $\sqrt {\psi _1
}$, $\sqrt {\psi _2 }$.

Из определения параметров видим, что
$$
e'_ +  e'_ -   = e''_ +  e''_ -   =  - (r_1  + r_2 )^2
$$
отрицательно на $\delta_1$, $\delta_2$ и положительно на $\delta_3$.
Сумма корней для системы $\delta_3$
$$
\begin{array}{l}
e'_ +   + e'_ -   = 2s\displaystyle{ \frac{r_1  + r_2 } {r_1 }} <
0,\qquad  e''_ +   + e''_ -   = 2s \displaystyle{\frac{r_1  + r_2 }
{r_2 }} > 0.
\end{array}
$$
Получаем, что на $\delta _1$, $\delta _2$ положительная переменная
$\eta$ осциллирует между корнем $\psi _1$ и корнем $\psi _2$. Выбирая подходящим образом соответствие знаков $ \pm $ в
определении (\ref{neq3_4}), получим, что к первой группе надо
отнести $u_1 ,u_3$, а ко второй  --- $u_2 ,u_4$.
Группируя столбцы соответственно, получим матрицу отображения
\eqref{neq3_5} в виде
$$
\begin{array}{c||c  c| c  c||}
\multicolumn{1}{c}{} & \multicolumn{1}{c}{v_1 } & \multicolumn{1}{c}{v_2 }
& \multicolumn{1}{c}{w_1 } & \multicolumn{1}{c}{w_2 }  \\
{z_1 } & 1 & {} & 1 & {}  \\
{z_2 } & {} & 1 & {} & 1 \\
\end{array}
$$
В терминах теоремы~\ref{theo3} здесь $P = Q = 2$.
Следовательно интегральное многообразие $\mathcal{F}_{s}$ состоит из
одной компоненты, то есть каждой точке бифуркационных кривых $\delta
_1$, $\delta _2$ в фазовом пространстве отвечает одно периодическое
решение.

На кривой $\delta_3$ переменная $\eta $ изменяется между $e''_+$ и
$e''_-$. Поэтому распределение аргументов по группам таково: первая
группа состоит из $u_1 ,u_2$, а вторая --- из $u_3 ,u_4$.
Матрица \eqref{neq3_5} принимает вид
$$
\begin{array}{c||c  c| c  c||}
\multicolumn{1}{c}{} & \multicolumn{1}{c}{v_1 } & \multicolumn{1}{c}{v_2 }
& \multicolumn{1}{c}{w_1 } & \multicolumn{1}{c}{w_2 }  \\
{z_1 } & 1 & 1 & {} & {}  \\
{z_2 } & {} & {} & 1 & 1 \\
\end{array}
$$
Поэтому $P = 1,Q = 0$. Количество компонент $2^{P - Q}  = 2$.
Следовательно, точкам бифуркационной кривой $\delta _3$
соответствует два периодических решения.

\begin{figure}[ht]
\centering
\includegraphics[width=100mm,keepaspectratio]{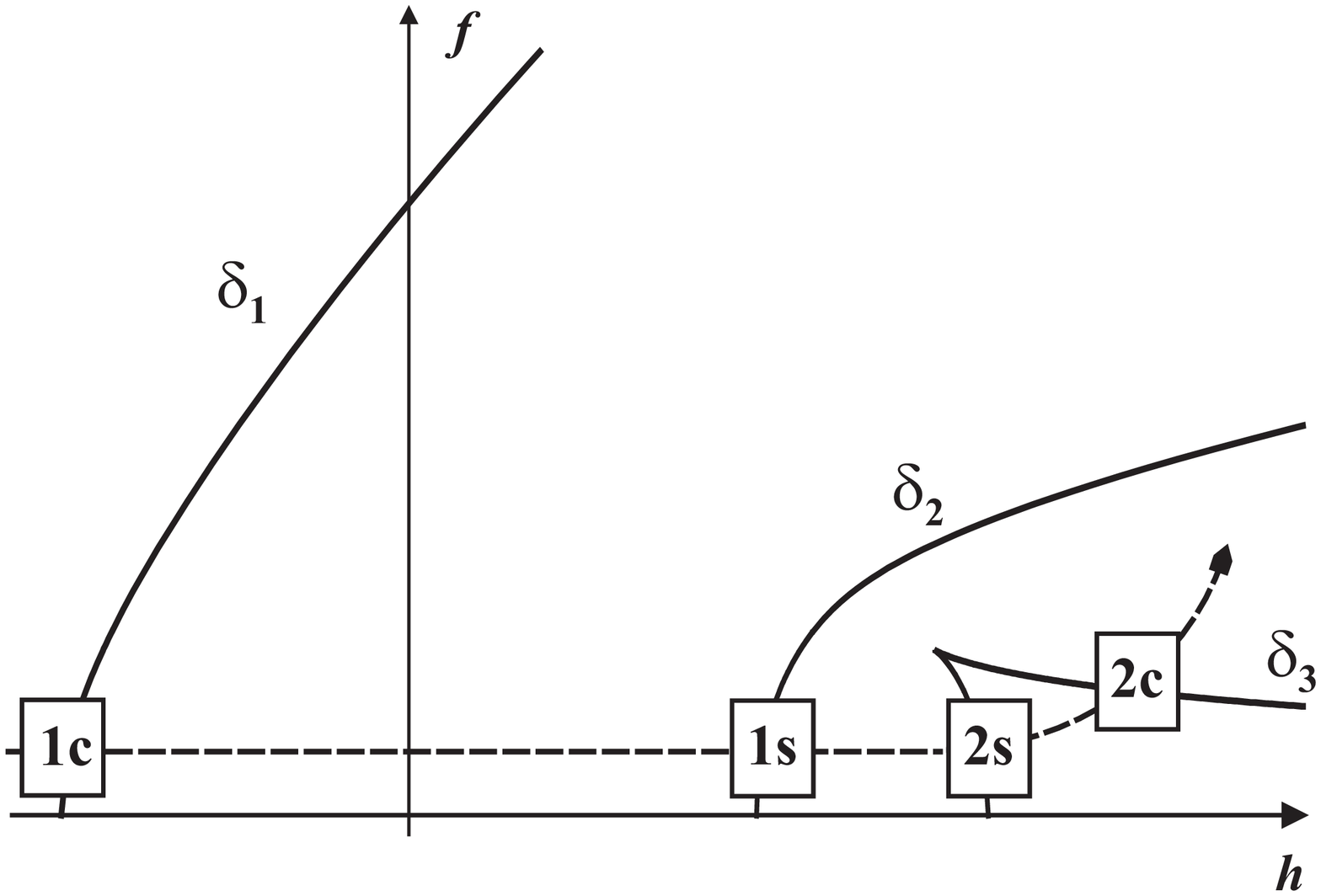}
\caption{Диаграмма первой критической подсистемы.}\label{fig_bogo}
\end{figure}

Несмотря на то, что общего аналитического решения для первой
критической подсистемы (случая Богоявленского) пока не получено,
теперь весьма просто установить всю фазовую топологию. Для этого
достаточно заметить, что на многообразии $\mathfrak{M}$
рассмотренные здесь периодические решения при заданном $s$ образуют
множество критических точек первого интеграла $F^2-2\tau H$, где
$\tau$ --- константа, приведенная в \eqref{neq3_1}. Характеристическое
уравнение симплектического оператора, порожденного этим интегралом
на $\mathfrak{M}$, записывается так \cite{KhRyVU}
$$
\mu^2 - \ds{\frac{2}{s}(r_1+r_2)^2[r_1^2 r_2^2+(r_1+r_2)^2 s^2]}=0.
$$
Оно, очевидно, имеет пару мнимых корней на $\delta_1$ и на ветвях
$\delta_3$, идущих от точек возврата в бесконечность, и пару
вещественных корней (разного знака) на $\delta_2$ и между точками
возврата $\delta_3$. Соответственно, в первом случае решения состоят
из критических точек ранга 1 типа <<центр>>, а во втором --- из
критических точек ранга 1 типа <<седло>>. В частности, в точках
возврата $\mu^2=0$, так что критические точки вырождены. Эти утверждения можно считать
технически доказанными в работах \cite{ZotRCD,ZotRus} при исследовании
боттовости интеграла $F$, но характеристических уравнений и чисел там не
приводится. На рис.~\ref{fig_bogo} представлена бифуркационная
диаграмма, состоящая из кривых \eqref{neq3_1}, с указанием типов
критических точек и количества критических окружностей.
Зная количество периодических решений, сразу же выписываем
последовательность бифуркаций вдоль пунктирной стрелки:
$$
\begin{array}{c}
\varnothing \to S^1 \to \bT^2 \to (S^1 \vee S^1){\times}S^1 \to
2\bT^2 \to 2(S^1 \vee S^1){\times}S^1 \to 4\bT^2 \to 2 \bT^2 \cup 2
S^1 \to 2\bT^2 .
\end{array}
$$

Полученные результаты совпадают с выводами работы \cite{ZotRCD}, но
здесь они доказаны очень просто и без привлечения плохо
формализуемых рассуждений. Справедливости ради заметим, что для
достижения этой простоты понадобилось 10 лет после выхода статьи
\cite{ZotRCD}.

\section{Вторая критическая система}\label{sec4}
На многообразии $\mathfrak{N}$, заданном уравнениями (\ref{neq2_2}),
(\ref{neq2_5}) рассматриваем интегральное отображение
\begin{equation}\label{neq4_1}
\mF = L \times M: \mathfrak{N} \to {\bf{R}}^2,
\end{equation}
порожденное интегралами (\ref{neq2_6}). Соответственно набор
параметров ${\bf{f}}$ есть точка плоскости ${\bf{f}} = (\ell,m)$.
Вводя вспомогательные переменные $s_1,s_2$, обозначим
\begin{equation}\notag
\begin{array}{l}
\displaystyle{\Psi (s_1 ,s_2 ) = 4ms_1 s_2  - 2\ell (s_1  + s_2 ) +
\frac{1}{m}(\ell ^2  - 1)},  \qquad \Phi (s)=\Psi(s,s).
\end{array}
\end{equation}
Тогда алгебраическое решение имеет вид \cite{KhSavDan, KhSav}
\begin{equation}\label{neq4_2}
\begin{array}{l}
\displaystyle{\alpha _1  = \frac{\mathstrut 1} {{2(s_1  - s_2 )^2
}}[(s_1 s_2
- a^2 )\Psi(s_1,s_2) + S_1 S_2 \varphi _1 \varphi _2 ], }\\
\displaystyle{\alpha _2  = \frac{\mathstrut 1} {{2(s_1  - s_2 )^2
}}[(s_1 s_2
- a^2)\varphi _1 \varphi _2  -  \Psi(s_1,s_2) S_1 S_2], }\\
\displaystyle{\beta _1  =  - \frac{\mathstrut 1} {{2(s_1  - s_2 )^2
}}[(s_1
s_2  - b^2)\varphi _1 \varphi _2  - \Psi(s_1,s_2) S_1 S_2 ], }\\
\displaystyle{\beta _2  = \frac{\mathstrut 1} {{2(s_1  - s_2 )^2
}}[(s_1 s_2
- b^2 )\Psi(s_1,s_2) + S_1 S_2 \varphi _1 \varphi _2 ], }\\
\displaystyle{\alpha _3  = \frac{\mathstrut r} {{s_1  - s_2 }}S_1
,\quad
\beta _3  =\frac{r} {{s_1  - s_2 }}S_2, }\\
\displaystyle{\omega _1  = \frac{\mathstrut r} {{2(s_1  - s_2
)}}(\ell - 2ms_1 )\varphi _2,\quad \omega _2  = \frac{r} {{2(s_1 -
s_2
)}}(\ell  - 2ms_2)\varphi _1, }\\
\displaystyle{\omega _3  = -\frac{\mathstrut 1} {{s_1  - s_2 }}(S_2
\varphi _1 + S_1 \varphi _2 )}.
\end{array}
\end{equation}
Здесь
\begin{equation}\label{neq4_3}
\displaystyle{S_1 = \sqrt {\mstrut s_1^2 - a^2},} \qquad
\displaystyle{\varphi_1 = \sqrt {\mstrut - \Phi (s_1)},} \qquad
\displaystyle{S_2 = \sqrt {\mstrut b^2 - s_2^2},} \qquad
\displaystyle{\varphi_2 = \sqrt {\mstrut \Phi (s_2)}.}
\end{equation}
Зависимость $s_1 ,s_2 $ от времени задана уравнениями
\begin{equation}\label{neq4_3a}
\dot {s}_1 = \frac{1} {2}\sqrt {\mstrut (a^2  - s_1^2 )\Phi (s_1 )}
,\quad \dot {s}_2 = \frac{1} {2}\sqrt {\mstrut (b^2  - s_2^2 )\Phi
(s_2 )}.
\end{equation}

По теореме~\ref{theo4} в качестве базисных радикалов можно сразу
принять выражения (\ref{neq4_3}). Максимальный многочлен имеет вид
$V(s) = (s^2-a^2)(s^2-b^2) \Phi(s)$. Его дискриминантное множество
состоит из прямых
\begin{equation}\label{neq4_4}
\ell  =  - 2ma \pm 1,\;\ell  = 2ma \pm 1,\;\ell  =  - 2mb \pm
1,\;\ell  = 2mb \pm 1
\end{equation}
и образует разделяющее множество на плоскости параметров $(\ell,m)$.

Для описания зависимости \eqref{neq4_2} вводим булевы аргументы
\begin{equation}\label{neq4_5}
\begin{array}{l}
u_1  = \lsgn S_1 ,\qquad u_2  = \lsgn \varphi_1 , \qquad
u_3  = \lsgn S_2 ,\qquad u_4  = \lsgn \varphi_2  \\
\end{array}
\end{equation}
и компоненты булевой вектор-функции $C$ (в скобках указаны
переменные, содержащие соответствующие мономы)
\begin{equation}\label{neq4_6}
\begin{array}{l l}
z_1  = u_1  \oplus u_2  \oplus u_3  \oplus u_4 & (\alpha_1, \beta_2);\\
z_2  = u_1  \oplus u_3, \quad z_3  = u_2  \oplus u_4 & (\alpha_2, \beta_1);\\
z_4  = u_1 & (\alpha_3);\\
z_5  = u_2 & (\omega_2);\\
z_6  = u_3 & (\beta_3);\\
z_7  = u_4 & (\omega_1);\\
z_8  = u_2 \oplus u_3 & (\omega_3);\\
z_9  = u_1 \oplus u_4  & (\omega_2).
\end{array}
\end{equation}

\begin{figure}[ht]
\centering
\includegraphics[width=65mm,keepaspectratio]{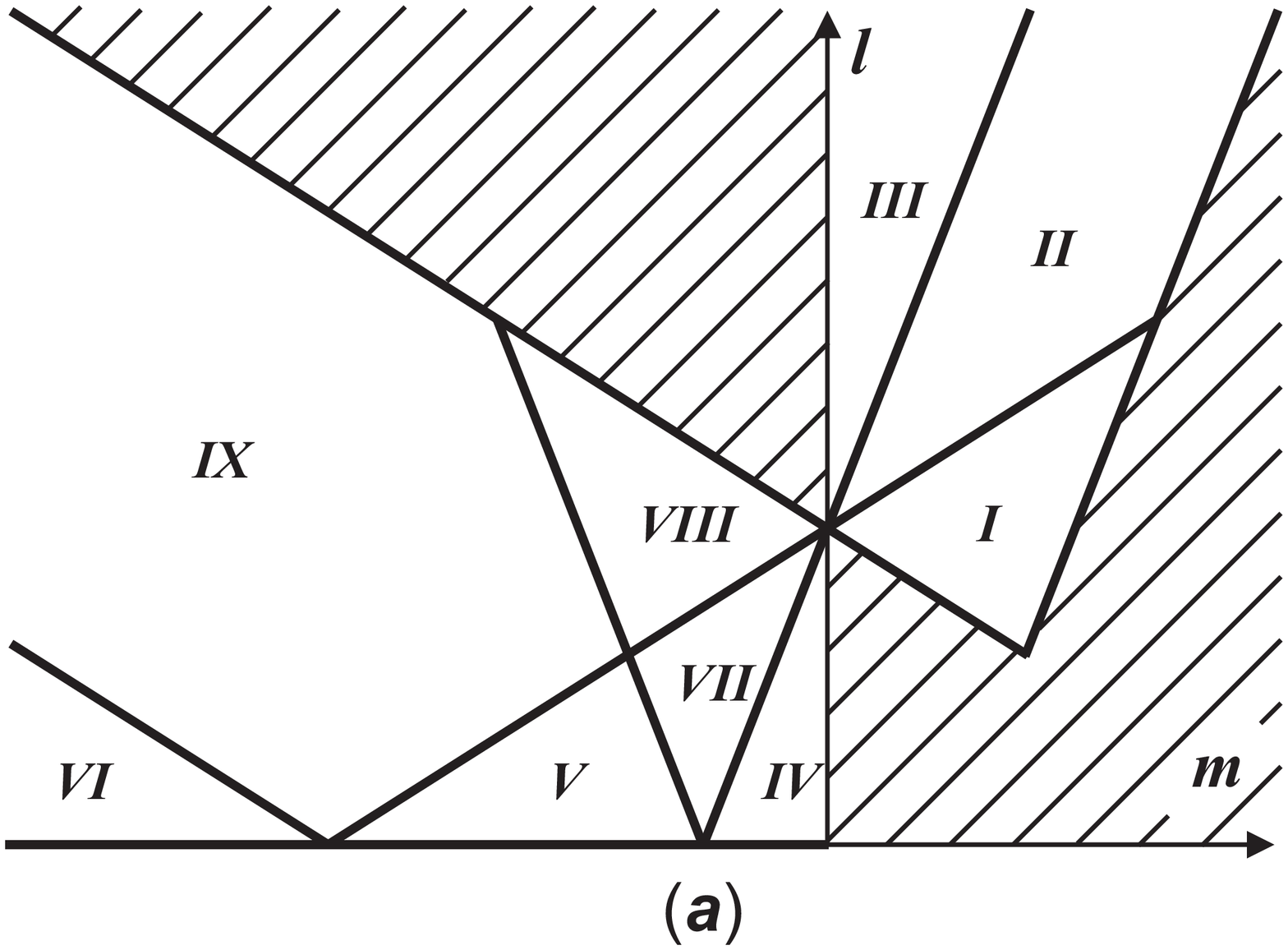}\hspace{10mm}
\includegraphics[width=65mm,keepaspectratio]{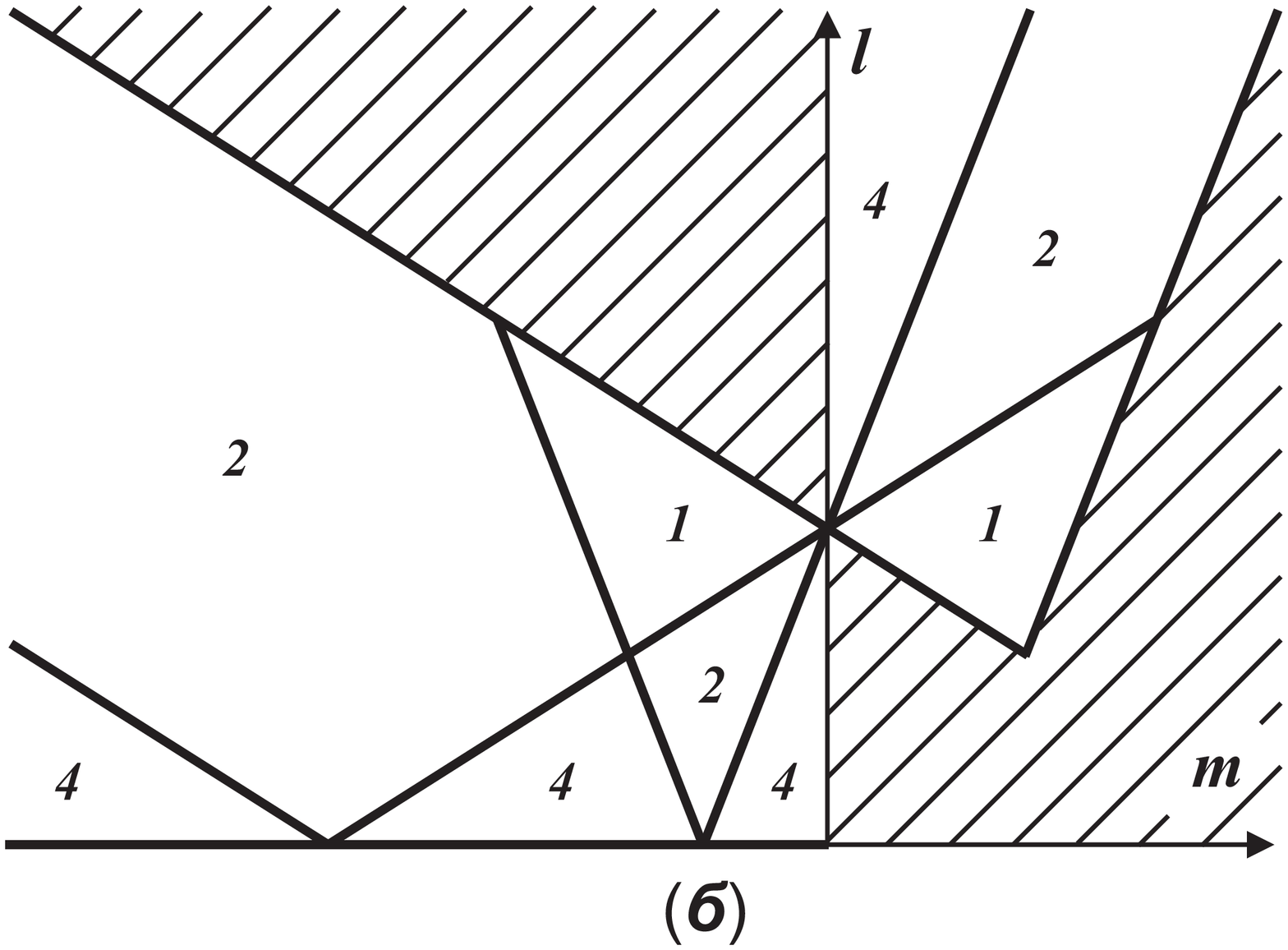}
\caption{Разделяющие кривые, кодировка областей и количество
торов.}\label{fig_sav_bif}
\end{figure}

Для определения допустимой области на плоскости $(\ell,m)$ заменим
радикалы в \eqref{neq4_5} на их квадраты, тогда в соответствии с \eqref{neq1_9} необходимое и достаточное условие вещественности решения
примет вид $C({\bf{u}}) = 000000000$, что равносильно ${\bf{u}} = 0000$, то есть
\begin{equation}\notag
S_1^2  \geqslant 0, \qquad S_2^2  \geqslant 0, \qquad  \varphi _1^2
\geqslant 0, \qquad \varphi _2^2 \geqslant 0.
\end{equation}
Очевидно это условие сильнее, чем условие вещественности решений \eqref{neq4_3a}. Полагая в силу симметрии $\ell \geqslant 0$, найдем
допустимую область
\begin{equation}\notag
\begin{array}{l}
\ell \geqslant \max (2ma - 1, - 2mb + 1),\quad m > 0;  \\
\ell \leqslant  - 2mb + 1,\quad m < 0;  \\
\ell  = 1,\quad m = 0.
\end{array}
\end{equation}
Это --- незаштрихованная часть на рис.~\ref{fig_sav_bif},{\it а}.
Здесь же занумерованы области регулярности \ts{I}~--~\ts{IX}.

Для установления количества компонент интегральных многообразий
заполним таблицу достижимых областей для переменных $s_1 ,s_2 $.
Обозначим корни $\Phi $ через
\begin{equation}\notag
\tau _1  = \frac{{\ell  - {\mathop{\rm sgn}} m}}{{2m}} < \tau _2  =
\frac{{\ell  + {\mathop{\rm sgn}} m}}{{2m}}.
\end{equation}
Полная информация представлена в табл.~1.

\begin{remark}\label{rem81} Напомним договоренность $\cite{KhNDnew}$: для случая, когда некоторая переменная
может пересекать бесконечность в конечные моменты времени, будем
использовать следующее обозначение --- запись $x\in [\,A(\pm
\infty)B\,]$ означает, что $A>B$, а переменная $(x -(A+B)/2)^{-1}$
осциллирует на отрезке $[\, -((A-B)/2)^{-1}, ((A-B)/2)^{-1}\,]$.
\end{remark}

\begin{table}[ht]
\centering
\begin{tabular}{|c| c| c| c| c|}
\multicolumn{5}{r}{\fts{Таблица 1}}\\
\hline
\begin{tabular}{c}Номер\\области\end{tabular}&\begin{tabular}{c}Корни\end{tabular}&\begin{tabular}{c}Область\\изменения $s_1$\end{tabular}
&\begin{tabular}{c}Область\\изменения $s_2$\end{tabular}&\begin{tabular}{c}Первая\\группа\end{tabular}\\
\hline
\ts{I}&$ - b < \tau _1  < b < a < \tau _2 $&$[\,a,\tau _2 \,]$&$[\, - b,\tau _1 \,]$&--\\
\hline
\ts{II}&$b < \tau _1  < a < \tau _2 $&$[\,a,\tau _2 \,]$&$[\, - b,b\,]$&4\\
\hline
\ts{III}&$a < \tau _1  < \tau _2 $&$[\,\tau _1 ,\tau _2 \,]$&$[\, - b,b\,]$&14\\
\hline
\ts{IV}&$\tau _1  <  - a < a < \tau _2 $&$[\,\tau _2 \,(\pm \infty)\, \tau _1 \,]$&$[\, - b,b\,]$&14\\
\hline
\ts{V}&$ - a < \tau _1  <  - b < b < \tau _2  < a$&$[\,a \,(\pm \infty)\, -a\,]$&$[\, - b,b\,]$&24\\
\hline
\ts{VI}&$ - a <  - b < \tau _1  < \tau _2  < b < a$&$[\,a \,(\pm \infty)\, -a\,]$&$[\,\tau _1 ,\tau _2 \,]$&23\\
\hline
\ts{VII}&$\tau _1  <  - a <  - b < b < \tau _2  < a$&$[\,a \,(\pm \infty)\, \tau _1 \,]$&$[\, - b,b\,]$&4\\
\hline
\ts{VIII}&$\tau _1  <  - a <  - b < \tau _2  < b < a$&$[\,a \,(\pm \infty)\, \tau _1 \,]$&$[\, - b,\tau _2 \,]$&--\\
\hline
\ts{IX}&$ - a < \tau _1  <  - b < \tau _2  < b < a$&$[\,a \,(\pm \infty)\, -a\,]$&$[\, - b,\tau _2 \,]$&2\\
\hline
\end{tabular}
\end{table}

Решая задачу определения количества связных компонент, в матрице
функции \eqref{neq4_6} по теореме~\ref{theo2} можно отбросить все
строки, кроме отвечающих $z_4 ,\ldots, z_7$. Остается единичная матрица.
Из нее можно вычеркивать столбцы, соответствующие второй группе, и
те строки, где в этих столбцах стоит 1. В результате остается
единичная матрица размерности $P$, равной количеству аргументов
\mbox{первой} группы. Если таковых нет, то $P=0$. Применяя
теорему~\ref{theo3}, получим следующее утверждение о количестве
связных компонент (торов Лиувилля) в составе $\mF_{\ell,m}$.

\begin{theorem}\label{th4}
В соответствии с нумерацией областей на рис.~\ref{fig_sav_bif},{\it
а} регулярные интегральные многообразия второй критической
подсистемы таковы: а) $\mbf{T}^2$ в областях \ts{I}, \ts{VIII}; б)
$2\mbf{T}^2$ в областях \ts{II}, \ts{VII}, \ts{IX}; в) $4\mbf{T}^2$
в областях \ts{III}, \ts{IV}, \ts{V}, \ts{VI}.
\end{theorem}
Число торов Лиувилля для наглядности показано на
рис.~\ref{fig_sav_bif},{\it б}.

К критическим случаям --- прямым (\ref{neq4_4}) --- применима та же
функция, и в результате действий с ее матрицей получаем следующие
общие закономерности:

1) если при переходе через отрезок бифуркационной прямой
(\ref{neq4_4}) число торов Лиувилля изменяется с $n$ на 0 $(n =
1,2)$, то критическая поверхность имеет $n$ компонент связности и,
следовательно, диффеоморфна $n S^1$;

2) если при таком переходе число торов меняется с $n$ на $2n$ $(n =
1,2)$, то критическая поверхность имеет $n$ компонент связности и,
следовательно, диффеоморфна $n(S^1\vee S^1){\times}S^1$.

Отметим, что переход через ось $O\ell$ невозможен, так как при $m =
0$ обязательно $\ell  = 1$.

Приведенные выше утверждения не относятся к полупрямой
\begin{equation}\label{neq4_7}
\ell  = 0, \qquad m \leqslant 0,
\end{equation}
так как здесь, строго говоря, критических точек отображения
(\ref{neq4_1}) не возникает, а возникают как особенность
индуцированной симплектической структуры \cite{KhSav}, так и
нарушения гладкости фазового пространства $\mathfrak{N}$. Поэтому
разберем этот случай особо, попутно применяя технику избавления от
особенностей в радикалах, связанных с уходом переменных на
бесконечность (в регулярном случае это не оговаривается, поскольку
принципиально ничего не меняет). Имеем при условии (\ref{neq4_7})
\begin{equation}\label{neq4_8}
s_1  \in ( - \infty , - s_1^* ] \cup [s_1^* , + \infty ), \qquad s_2
\in [ - s_2^* ,s_2^* ],
\end{equation}
где
\begin{equation}\notag
s_1^*  = \max \{ a,c\}, \qquad s_2^*  = \min \{ b,c\}, \qquad c =  -
\frac{1}{{2m}}> 0.
\end{equation}
Выражения под радикалами (\ref{neq4_3}), а также максимальный
многочлен становятся четными функциями от $s_1 ,s_2 $, что порождает
дополнительную симметрию. Чтобы избавиться от лишнего ветвления,
нужно ввести переменные
\begin{equation}\notag
\xi _1  = s_1^2 , \qquad \xi _2  = s_2^2
\end{equation}
но тогда к базисным радикалам (\ref{neq4_3}) добавятся и $\sqrt {\xi
_1 }$, $\sqrt {\xi _2 }$. Теперь для того чтобы при <<отражении>>
$\xi _1$ от $ + \infty$ менялся знак только одного радикала,
переопределим базисные радикалы так
\begin{equation}\notag
\begin{array}{lll}
R_1  = \sqrt {\ds{\frac{{\xi _1  - a^2 }}{{\xi _1 }}}}, & R_2 =
\sqrt {\ds{\frac{{\xi _1  - c^2 }}{{\xi _1 }}}}, &
R_3  = \ds{\frac{1}{{\sqrt {\xi _1 } }}},\\[4mm]
R_4  = \sqrt {\mstrut b^2  - \xi _2 },& R_5  = \sqrt {\mstrut c^2 -
\xi _2 },&  R_6  = \sqrt {\mstrut \xi _2 },
\end{array}
\end{equation}
и положим $u_\gamma   = \lsgn  R_\gamma$ $(\gamma = 1 ,\ldots, 6)$.
Зависимости (\ref{neq4_2}) представим в виде:
\begin{equation}\notag
\begin{array}{lll}
\alpha _1  = \ds{\frac{1}{{2\varkappa ^2 }}}(\varkappa_1 \theta   +
R_1 R_2 R_4 R_5 ), &
\alpha _2  = \ds{\frac{1}{{2\varkappa ^2 }}}(\varkappa _1 R_2 R_5  - R_1 R_4 \theta) ,   & \alpha _3  = \ds{\frac{r}{\varkappa }}R_1,\\[4mm]
\beta _1  = \ds{\frac{1}{{2\varkappa ^2 }}}(\varkappa _2 R_2 R_5  -
R_1 R_4 \theta) , &
\beta _2  = \ds{\frac{1}{{2\varkappa ^2 }}}(\varkappa _2 \theta  + R_1 R_2 R_4 R_5) , & \beta _3  = \ds{\frac{r}{\varkappa }}R_3 R_4 ,\\[4mm]
\omega _1  =  - \ds{\frac{{m\, r}}{\varkappa }}R_5 , & \omega _2  =
- \ds{\frac{{m\,r}}{\varkappa}} R_4 R_6 , & \omega _3  =
-\ds{\frac{1}{\varkappa }}(R_2 R_4  + R_1 R_5 ),
\end{array}
\end{equation}
где $\varkappa  = 1 - R_3 R_6$, $\varkappa _1 = R_6 - a^2 R_3$,
$\varkappa _2  = R_6  - b^2 R_3$, $\theta = 4 m R_6  - m^{-1} R_3$.
В полученных выражениях фазовых переменных после приведения их к
многочленам от базисных радикалов встречается 12 различных мономов,
которые порождают булеву вектор-функцию $ C:\mB^6  \to \mB^{12}$ по
формулам (в последнем столбце в скобках указаны переменные,
содержащие такие комбинации):
\begin{center}
\begin{tabular}{l l m{3cm}}
$z_1  = u_3  \oplus u_6,$ & $z_2  = u_1  \oplus u_2  \oplus u_4  \oplus u_5$ &  $(\alpha _1 ,\beta _2 );$\\
$z_3  = u_1  \oplus u_3  \oplus u_4,$ & $z_4  = u_1  \oplus u_4  \oplus u_6 $ & \multirow{2}{3cm}{$(\alpha _2 ,\beta _1);$}\\
$z_5  = u_2  \oplus u_3  \oplus u_5,$ & $z_6  = u_2  \oplus u_5
\oplus u_6$ & {}\\
$z_7  = u_1$ &{}& $(\alpha _3 );$\\
$z_8= u_3  \oplus u_4$ & {}& $(\beta _3 );$\\
$z_9 = u_5$ & {}& $(\omega _1 );$\\
$z_{10} = u_2  \oplus u_6$ & {}& $(\omega _2 );$\\
$z_{11} = u_2  \oplus u_4 ,$ & $z_{12}  = u_1  \oplus u_5$ &
$(\omega _3 ).$
\end{tabular}
\end{center}

Дальнейшие эквивалентные операции над матрицей таковы: с помощью
$z_7 ,z_9 $ обнуляем остальные элементы столбцов $u_1 ,u_5 $. После
этого отбрасываем зависимые компоненты $z_5  = z_2  \oplus z_3$,
$z_6 = z_1  \oplus z_2  \oplus z_3$, $z_8  = z_3$, $z_{10}  = z_6$,
$z_{11}  = z_2$, $z_{12}  \equiv 0$. Далее учтем, что аргументы
$u_3,u_6$ всегда относятся ко второй группе. Преобразуем $z_3
\mapsto z_3 \oplus z_1$, исключим $(z_1 ,u_3 )$. После этого
оказывается $z_4  = z_3$, исключим $z_4$. В столбце $u_6$ осталась
одна единица в строке $z_3$, исключим $(z_3 ,u_6 )$. Осталась
матрица $3{\times}4$, определяющая редуцированную функцию
$$
\begin{array}{l}
z_2  = u_2  \oplus u_4,\qquad z_7  = u_1,\qquad z_9  = u_5.
\end{array}
$$

Рассматриваем три участка полупрямой (\ref{neq4_7}), образованные
пересечениями с прямыми (\ref{neq4_4}):

$1^\circ)$ $0 > m >  - \ds{\frac{1}{{2a}}} \Rightarrow c > a$,
$s_1^* = c$, $s_2^*  = b$; \vspace{1mm}

$2^\circ)$ $ - \ds{\frac{1}{{2a}}} > m >  - \ds{\frac{1}{{2b}}}
\Rightarrow b < c < a$ , $s_1^*  = a$, $s_2^*  = b$;\vspace{1mm}

$3^\circ)$ $ - \ds{\frac{1}{{2b}}} > m \Rightarrow c < b$, $s_1^* =
a$, $s_2^*  = c$.\vspace{1mm}

В случае $(1^\circ)$ к первой группе относятся $u_1 ,u_5 $, в случае
$(2^\circ)$ --- $u_2 ,u_5 $, а в случае $(3^\circ)$ --- $u_2 ,u_4 $. В
терминах теоремы~\ref{theo3} имеем: $(1^\circ)$ $P = 2,Q = 0$;
$(2^\circ)$ $P = 2,Q = 1$; $(3^\circ)$ $P = 1,Q = 0$. Получаем
доказательство следующего утверждения, которое нельзя установить
никаким локальным анализом невырожденных особенностей.
\begin{theorem}\label{theo5}
Интегральные поверхности на участках прямой $\ell  = 0$, примыкающих
к областям \ts{IV}\,--\,\ts{VI}, таковы: $4\mbf{T}^2 $ для области
\ts{IV}, $2\mbf{T}^2 $ для областей \ts{V}, \ts{VI}.
\end{theorem}
В рамках гладкой теории первая из таких бифуркаций (выход на границу
из области \ts{IV} с сохранением числа торов) была бы невозможна.
Возможно, что своим существованием эти бифуркации обязаны особенностям фазового
пространства.

\section{Третья критическая система}\label{sec5}
На многообразии $\mathfrak{O}$, заданном уравнениями (\ref{neq2_2}),
(\ref{neq2_7}), рассматриваем интегральное отображение
\begin{equation}\label{neq5_1}
\mF = S \times \mathrm{T}: \mathfrak{O} \to {\bf{R}}^2,
\end{equation}
порожденное интегралами (\ref{neq2_8}). Соответственно набор
параметров ${\bf{f}}$ есть точка плоскости ${\bf{f}} = (s,\tau)$.
Введем также обозначения
\begin{equation}\label{neq5_2}
\begin{array}{l}
\sigma = \tau^2-2p^2 \tau+r^4, \quad
\chi=\sqrt{\ds{\frac{4s^2\tau+\sigma}{4s^2}}}, \quad
\varkappa=\sqrt{\sigma}.
\end{array}
\end{equation}
Если $\chi, \varkappa$ вещественные, то считаем их положительными,
если чисто мнимые, то считаем положительной мнимую часть.

Обозначим пару вспомогательных переменных через $t_1,t_2$. Введем
обозначения алгебраических радикалов
\begin{equation}\label{neq5_3}
\begin{array}{ll}
K_1=\sqrt{\mstrut t_1+\varkappa}\,, & K_2=\sqrt{\mstrut t_2+\varkappa}\,,\\
L_1=\sqrt{\mstrut t_1-\varkappa}\,, & L_2=\sqrt{\mstrut t_2-\varkappa}\,,\\
M_1=\sqrt{\mstrut t_1+\tau+r^2}\,, & M_2=\sqrt{\mstrut
t_2+\tau+r^2}\,,
\\[1.5mm]
N_1=\sqrt{\mstrut t_1+\tau-r^2}\,, &
N_2=\sqrt{\mstrut t_2+\tau-r^2}\,,\\
V_1 = \sqrt{\mstrut \ds{\frac{t_1^2-4s^2\chi^2}{s\tau}}}\,, & V_2 =
\sqrt{\mstrut \ds{\frac{t_2^2-4s^2\chi^2}{s\tau}}}
\end{array}
\end{equation}
и пусть
\begin{equation}\notag
\begin{array}{l}
U_1=K_1 L_1, \quad U_2=K_2 L_2,\quad R=K_1 K_2 + L_1 L_2,\\
\mathcal{A}=[(t_1+\tau+r^2)(t_2+\tau+r^2)-2(p^2+r^2)r^2]\tau, \\
\mathcal{B}=[(t_1+\tau-r^2)(t_2+\tau-r^2)+2(p^2-r^2)r^2]\tau.
\end{array}
\end{equation}
Тогда на фиксированном интегральном многообразии
\begin{equation}\notag
\mF_\mathbf{f} =\{S=s,\mathrm{T}=\tau\} \subset \mathfrak{O}, \qquad
\mathbf{f}=(s,\tau)
\end{equation}
будем иметь \cite{KhRCD09}
\begin{equation}\label{neq5_4}
\begin{array}{l}
\displaystyle{\alpha_1=(U_1-U_2)^2\frac{(\mathcal{A}-r^2 U_1 U_2)(4
s^2 \tau+U_1 U_2)-(\tau+r^2) s\tau M_1 N_1 V_1 M_2 N_2 V_2}{4 r^2
s\, \tau (t_1^2-t_2^2)^2},
} \\[3mm]
\displaystyle{\alpha_2=\ri (U_1-U_2)^2 \frac{(\mathcal{A} -r^2 U_1
U_2)s\tau V_1 V_2 -(4 s^2 \tau+U_1 U_2)(\tau+r^2) M_1 N_1 M_2 N_2}{4
r^2 s\, \tau (t_1^2-t_2^2)^2},
}\\[3mm]
\displaystyle{\alpha_3= \frac{R }{2 r } \, \frac {M_1
M_2}{t_1+t_2},} \\
\displaystyle{\beta_1=\ri (U_1-U_2)^2 \frac{(\mathcal{B}+r^2 U_1
U_2)s\tau  V_1 V_2-(4 s^2 \tau+U_1 U_2)(\tau-r^2) M_1 N_1 M_2 N_2}{4
r^2 s\, \tau (t_1^2-t_2^2)^2},
} \\[3mm]
\displaystyle{\beta_2=-(U_1-U_2)^2 \frac{(\mathcal{B} + r^2 U_1
U_2)(4 s^2 \tau + U_1 U_2)-(\tau-r^2) s\tau  M_1 N_1 V_1 M_2 N_2
V_2}{4 r^2 s\, \tau (t_1^2-t_2^2)^2},
}\\[3mm]
\displaystyle{\beta_3= - \ri \frac{R}{2 r} \, \frac { N_1
N_2}{t_1+t_2},}
\\[3mm]
\displaystyle{\omega_1=  \ri \frac{R }{4 r s\,
\sqrt{2}} \, \frac{ U_1 N_1 M_2 V_2 - M_1 V_1 U_2 N_2}{ t_1^2-t_2^2},} \\[3mm]
\displaystyle{\omega_2= \frac{R}{4 r s\, \sqrt{2} } \,
\frac{U_1 M_1 N_2 V_2 - N_1 V_1 U_2 M_2}{t_1^2-t_2^2},} \\[3mm]
\displaystyle{\omega_3=  - \ri \frac{{U_1-U_2}}{\sqrt{2}}\frac{V_1
M_2 N_2 + M_1 N_1 V_2}{t_1^2-t_2^2}.}
\end{array}
\end{equation}
При этом переменные $t_1,t_2$ удовлетворяют дифференциальным
уравнениям
\begin{equation}\label{neq5_5}
\begin{array}{ll}
\displaystyle{(t_1-t_2) \dot t _1  = \sqrt{\frac{1}{2 s
\tau}(t_1^2-4s^2\chi^2)(t_1^2-\sigma)[(t_1+\tau)^2-r^4]}\, ,} \\
[3mm] \displaystyle{(t_1-t_2) \dot t _2  = \sqrt{\frac{1}{2 s
\tau}(t_2^2-4s^2\chi^2)(t_2^2-\sigma)[(t_2+\tau)^2-r^4]}\,.}
\end{array}
\end{equation}

Максимальный многочлен имеет вид
\begin{equation}\notag
V(t)=\ds{\frac{1}{s\tau}}(t^2-4s^2\chi^2)(t^2-\sigma)[(t+\tau)^2-r^4],
\end{equation}
и его дискриминантное множество (разделяющее множество на плоскости
параметров $s,\tau$) состоит из прямых
\begin{equation}\notag
s=0, \qquad s=\pm a,\qquad s=\pm b, \qquad \tau=(a\pm b)^2, \qquad
\tau=0
\end{equation}
и кривой $\chi=0$ или, в явном виде,
\begin{equation}\label{neq5_6}
\tau^2+2(2s^2-p^2)\tau+r^4=0.
\end{equation}
Эта кривая, очевидно, состоит из простой замкнутой кривой, вписанной в
прямоугольник
\begin{equation}\notag
-b \leqslant s \leqslant b, \qquad (a-b)^2 \leqslant \tau \leqslant
(a+b)^2,
\end{equation}
и двух бесконечных ветвей, лежащих симметрично относительно оси
$O\tau$ в квадрантах $\tau<0, s \geqslant a$ и $\tau<0, s \leqslant
-a$.

\begin{remark}\label{rem91}
Отметим, что условие $\chi^2>0$ выполнено на плоскости $(s,\tau)$ в
единственной связной области, лежащей между замкнутой кривой и
бесконечными ветвями. Это упростит рассуждения ниже.
\end{remark}

По теореме~\ref{theo4} в качестве базисных радикалов можно взять 10
радикалов (\ref{neq5_3}). Для краткости некоторых формул будем
обозначать их также через $R_{i \gamma}$:
\begin{equation}\notag
R_{i 1}=K_i,\;R_{i 2}=L_i,\;R_{i 3}=M_i,\;R_{i 4}=N_i,\;R_{i 5}=V_i
\qquad  (i=1,2).
\end{equation}
Вводим булевы переменные
\begin{equation}\label{neq5_7}
u_\gamma = \lsgn R_{1 \gamma}^2,\; u_{5+\gamma} = \lsgn R_{2
\gamma}^2 \qquad (\gamma =1 ,\ldots, 5)
\end{equation}
для решения задачи определения достижимых областей и
\begin{equation}\notag
u_\gamma = \lsgn R_{1 \gamma},\; u_{5+\gamma} = \lsgn R_{2 \gamma}
\qquad (\gamma =1 ,\ldots, 5)
\end{equation}
для задачи исследования фазовой топологии.

Из (\ref{neq5_4}) найдем полный набор мономов, определяющих
алгебраическую структуру надстройки. Из уравнений для
${\bs \alpha}, {\bs \beta}$ получаем компоненты булевой
вектор-функции
\begin{equation}\notag
\begin{array}{lll}
z_1  = u_1  \oplus u_2  \oplus u_6  \oplus u_7   &{}& (K_1 L_1 K_2 L_2); \\
z_2  = u_3  \oplus u_4  \oplus u_5  \oplus u_8  \oplus u_9 \oplus u_{10} &{}& ( M_1 N_1 V_1 M_2 N_2 V_2); \\
z_3  = u_5  \oplus u_{10} &{}& ( V_1  V_2); \\
z_4  = u_1  \oplus u_2  \oplus u_5  \oplus u_6  \oplus u_7  \oplus u_{10} &{}& (K_1 L_1 V_1 K_2 L_2 V_2); \\
z_5  = u_3  \oplus u_4  \oplus u_8  \oplus u_9  &{}& (M_1 N_1 M_2 N_2); \\
z_6  = u_1  \oplus u_2  \oplus u_3  \oplus u_4  \oplus u_6  \oplus u_7  \oplus u_8  \oplus u_9 &{}& (K_1 L_1 M_1 N_1 K_2 L_2 M_2 N_2 ); \\
z_7  = u_1  \oplus u_3  \oplus u_6  \oplus u_8   &{}& (K_1 M_1 K_2 M_2 ); \\
z_8  = u_2  \oplus u_3  \oplus u_7  \oplus u_8   &{}& (L_1 M_1 L_2 M_2); \\
z_9  = u_1  \oplus u_4  \oplus u_6  \oplus u_9  &{}& (K_1 N_1 K_2 N_2); \\
z_{10}  = u_2  \oplus u_4  \oplus u_7  \oplus u_9  &{}& (L_1 N_1 L_2
N_2);\\
z_{11}  = u_1  \oplus u_2  \oplus u_3  \oplus u_4  \oplus u_5 \oplus
u_6  \oplus u_7  \oplus u_8  \oplus u_9 \oplus u_{10} &{}& (K_1 L_1
M_1 N_1 V_1 K_2 L_2 M_2 N_2 V_2).
\end{array}
\end{equation}
Здесь в последнем столбце в скобках записаны соответствующие мономы
от базисных радикалов (для первой задачи надо взять квадраты
радикалов). Выражения для ${\bs \omega}$ добавляют еще 12 компонент:
\begin{equation}\notag
\begin{array}{lll}
z_{12}  = u_1  \oplus u_3  \oplus u_5 \oplus u_7 \oplus u_9 &\hspace*{1mm} & (K_1 M_1 V_1 L_2 N_2) ;\\
z_{13}  = u_2 \oplus u_4  \oplus u_6  \oplus u_8  \oplus u_{10}
 &{}&
(L_1 N_1 K_2 M_2 V_2);\\
z_{14}  = u_1  \oplus u_4  \oplus u_7  \oplus u_8  \oplus u_{10}
 &{}&
(K_1 N_1 L_2 M_2 V_2);\\
z_{15}  = u_2  \oplus u_3  \oplus u_5  \oplus u_6  \oplus u_{9}
 &{}&
(L_1 M_1 V_1 K_2 N_2);\\
z_{16}  = u_1  \oplus u_4  \oplus u_5  \oplus u_7  \oplus u_{8}
 &{}&
(K_1 N_1 V_1 L_2 M_2) ;\\
z_{17}  = u_2  \oplus u_3  \oplus u_6  \oplus u_9  \oplus u_{10}
 &{}&
(L_1 M_1 K_2 N_2 V_2) ;\\
z_{18}  = u_1  \oplus u_3  \oplus u_7  \oplus u_9  \oplus u_{10}
 &{}&
(K_1 M_1 L_2 N_2 V_2) ;\\
z_{19}  = u_2  \oplus u_4  \oplus u_5  \oplus u_6  \oplus u_{8}
 &{}&
(L_1 N_1 V_1 K_2 M_2) ;\\
z_{20}  = u_1  \oplus u_2  \oplus u_5  \oplus u_8  \oplus u_{9}
 &{}&
(K_1 L_1 V_1 M_2 N_2) ;\\
z_{21}  = u_1  \oplus u_2  \oplus u_3  \oplus u_4  \oplus u_{10}
 &{}&
(K_1 L_1 M_1 N_1 V_2) ;\\
z_{22}  = u_5  \oplus u_6  \oplus u_7  \oplus u_8  \oplus u_{9}
 &{}&
(V_1 K_2 L_2 M_2 N_2) ;\\
z_{23}  = u_3  \oplus u_4  \oplus u_6  \oplus u_7  \oplus u_{10}
 &{}& (M_1 N_1 K_2 L_2 V_2).
\end{array}
\end{equation}

Вначале рассмотрим случай $\sigma >0$, в котором все подкоренные
выражения вещественны. Для вещественности значений (\ref{neq5_4})
необходимо и достаточно, чтобы в пространстве $\mB^{23}$ образом
построенного отображения $C: \mB^{10} \to \mB^{23}$ в трактовке
аргументов (\ref{neq5_7}) была точка
\begin{equation}\notag
\mbf{z}_0 = 00111100110111100001111.
\end{equation}
Запишем расширенную матрицу $\tilde C = ||C|\mbf{z}_0||$.
Элементарные преобразования ее строк и исключение нулевых строк не
влияют на результаты в обеих задачах. Поэтому поступим так. С
помощью строки $z_1$ сделаем единичным столбец $u_1$, прибавляя эту
строку (по модулю 2) ко всем строкам, кроме $z_1$, где в столбце
$u_1$ стоит 1. Далее с помощью строки $z_{10}$ сделаем единичным
столбец $u_2$, с помощью строки $z_{2}$ сделаем единичным столбец
$u_3$. Столбец $u_4$ не содержит единиц в неиспользованных строках.
Поэтому с помощью строки $z_{3}$ сделаем единичным столбец $u_5$ (то
есть 1 стоит только в строке $z_{3}$), а с помощью $z_{12}$
преобразуем столбец $u_6$. Остальные строки оказываются нулевыми.
Переобозначая компоненты редуцированной функции
$$
(z_1,z_{10},z_2,z_3,z_{12}) \to (z_1,z_2,z_3,z_4,z_5),
$$
получим матрицу $\tilde C$ в виде
\begin{equation}\label{neq5_8}
\begin{array}{c||c  c  c  c c c c c c c |c||}
\multicolumn{1}{c}{} & \multicolumn{1}{c}{u_1 } &
\multicolumn{1}{c}{u_2 } & \multicolumn{1}{c}{u_3 } &
\multicolumn{1}{c}{u_4 } & \multicolumn{1}{c}{u_5 } &
\multicolumn{1}{c}{u_6 } & \multicolumn{1}{c}{u_7 } &
\multicolumn{1}{c}{u_8 } & \multicolumn{1}{c}{u_9 } &
\multicolumn{1}{c}{u_{10} } & \multicolumn{1}{c}{\mbf{z}_0}
\\
{z_1 } & {1} & {} & {} & {1}   & {}  & {1} & {}  & {}  & {1} & {} & {1}\\
{z_2 } & {} & {1} & {} & {1}   & {}  & {}  & {1} & {}  & {1} & {} &  {1}\\
{z_3 } & {} & {} & {1} & {1}   & {}  & {}  & {}  & {1} & {1} & {} &  {1}\\
{z_4 } & {} & {} & {}  & {}    & {1} & {}  & {}  & {}  &  {} & {1} & {1}\\
{z_5 } & {} & {} & {}  & {}    & {}  & {1} & {1} & {1} & {1} & {1} & {0}\\
\end{array}.
\end{equation}
Итак, редуцированная функция $\hat{C}: \mB^{10} \to \mB^5$ такова:
\begin{equation}\label{neq5_9}
\begin{array}{ll}
z_{1}  = (u_1  \oplus u_6)  \oplus (u_4 \oplus u_9),\\
z_{2}  = (u_2  \oplus u_7)  \oplus (u_4 \oplus u_9),\\
z_{3}  = (u_3  \oplus u_8)  \oplus (u_4 \oplus u_9),\\
z_{4}  = u_5  \oplus u_{10},\\
z_{5}  = u_6  \oplus u_7  \oplus u_8 \oplus u_9 \oplus u_{10}.
\end{array}
\end{equation}
Это выражение окончательно в том смысле, что $\rk_2(\hat{C})=5$ и
количество компонент функции уменьшить в общем случае нельзя.
Условие вещественности переменных --- система вида \eqref{neq1_9} --- здесь запишется так
\begin{equation}\label{neq5_10}
\hat{C}\mbf{u}=11110.
\end{equation}
Равенство компонент $z_1 ,\ldots, z_4$ единице обеспечивает
вещественность переменных ${\bs \alpha}, {\bs \beta}$ (исходные
компоненты $z_1 ,\ldots, z_{11}$). Компонента $z_5$ --- это все, что
добавили выражения для $\bs \omega$ (все 12 исходных компонент
$z_{12} ,\ldots, z_{23}$). Интересно отметить, что условие $z_5=0$
выражает неотрицательность многочлена под радикалом второго
уравнения (\ref{neq5_5}) (фактически --- это знак максимального
многочлена от второй переменной разделения с учетом знака
произведения $s\,\tau$). Неотрицательность подкоренного выражения в
первом уравнении (\ref{neq5_5}) гарантирует следствие системы
\eqref{neq5_10}
\begin{equation}\notag
\begin{array}{l}
z_1 \oplus z_2 \oplus z_3 \oplus z_4 \oplus z_5 = u_1 \oplus u_2
\oplus u_3 \oplus u_4 \oplus u_5 =0.
\end{array}
\end{equation}

Система (\ref{neq5_10}) имеет $2^{10-5}=32$ решения, то есть для
нахождения достижимых областей в плоскости $(t_1,t_2)$ и,
соответственно, для установления допустимой области на плоскости
$(s,\tau)$ необходимо анализировать 32 системы неравенств.

Пусть $\sigma<0$. Тогда (см. по этому поводу замечание 1 работы
\cite{KhNDnew}) необходимо считать пары $K_1,L_1$ и $K_2,L_2$
комплексно сопряженными. Это добавляет в функцию $\hat{C}$ еще две
компоненты с условиями
\begin{equation}\label{neq5_11}
u_1 \oplus u_2=0, \qquad u_6 \oplus u_7=0,
\end{equation}
что увеличивает ранг отображения на единицу, и, соответственно,
уменьшает размерность пространства решений. Получаем 16 систем
неравенств. Это, по-прежнему, слишком много. Поэтому удобно
воспользоваться некоторыми сведениями о расположении корней. В
надежде, что повторное использование некоторых букв в локальном для
данного раздела смысле не повлечет недоразумений, обозначим
\begin{equation}\label{neq5_12}
m = -\tau - r^2, \qquad n = -\tau + r^2, \qquad v = 2|s|\chi
\end{equation}
и, соответственно, корни максимального многочлена
\begin{equation}\notag
e_1 = -\varkappa, \qquad e_2 = \varkappa, \qquad e_3 = m, \qquad e_4
= n, \qquad e_5 = -v, \qquad e_6 = v.
\end{equation}

Вначале заметим, что условие $z_4=1$ означает $V_1^2 V_2^2<0$. В
частности, отсюда сразу следует, что $\chi^2>0$. Напомним, что все
неравенства для подрадикальных выражений пишем как строгие,
рассматривая внутренние точки достижимых областей для регулярных
значений пары интегралов (то есть вне дискриминантного множества).
Условие $\chi^2>0$ по замечанию \ref{rem91} отсекает большое
количество областей как недопустимые. В силу произвола в нумерации
переменных разделения договоримся, что всегда будем считать во
внутренней точке достижимой области $t_1^2 < v^2, t_2^2 > v^2$. Это
соответствует выбору
\begin{equation}\label{neq5_13}
u_5= \neg \lsgn (s\tau),  \qquad u_{10}=\lsgn (s\tau).
\end{equation}
Добавим условия-импликации, вытекающие из очевидных неравенств
$e_1<e_2$ и $e_3<e_4$:
\begin{equation}\label{neq5_14}
(u_1 \to u_2) =1, \qquad (u_6 \to u_7)=1, \qquad (u_3 \to u_4) =1,
\qquad (u_8 \to u_9) = 1.
\end{equation}
Для набора $\mbf{u}$ в результате остается 10 значений --- по пять
систем неравенств для каждого из знаков произведения $s\tau$. Теперь
уже необходимо принять во внимание расположение корней в различных
случаях. Для этого, учитывая, что величины $\sigma,\varkappa,m,n$
зависят только от $\tau$, запишем
\begin{equation}\notag
\begin{array}{l}
\tau <0 \Rightarrow -\varkappa < m < n
<\varkappa;\\
\tau >0, \, \sigma >0 \Rightarrow \left[ \begin{array}{l} 0<\tau <(a-b)^2 \\
\tau>(a+b)^2
\end{array}\right.\; \Rightarrow m < n < - \varkappa.
\end{array}
\end{equation}
В первом случае добавляем условия-импликации
$$
(u_1 \to u_3)=1, \qquad (u_4 \to u_2)=1,\qquad (u_6 \to u_8)=1,
\qquad (u_9 \to u_7)=1
$$
и получаем ровно одно решение для каждого знака $s\tau$:
\begin{equation}\notag
\tau<0 \; \Rightarrow \begin{array}{l} \mbf{u}=0100101010 \quad (s\tau >0 )\\
\mbf{u}=0101001001 \quad (s\tau <0 )
\end{array}.
\end{equation}
Аналогично, во втором случае добавляем условия-импликации
$$
(u_4 \to u_1)=1, \qquad (u_9 \to u_6)=1
$$
и снова получаем ровно одно решение для каждого знака $s\tau$:
\begin{equation}\notag
\tau>0, \, \sigma>0  \; \Rightarrow \begin{array}{l} \mbf{u}=1101111000 \quad (s\tau >0 )\\
\mbf{u}=1100011011 \quad (s\tau <0 )
\end{array}.
\end{equation}
Из найденных решений сразу же выписываются промежутки для случая
$\sigma>0$:
\begin{equation}\label{neq5_15}
\begin{array}{lcl}
s<0, \,\tau<0 &  \Rightarrow & \left\{ \begin{array}{ll} \max\{-\varkappa,m,n\} \leqslant t_1 \leqslant \varkappa, & |t_1|\leqslant v \\
\max\{ -\varkappa,m \} \leqslant t_2 \leqslant \min\{ \varkappa,n
\}, & |t_2| \geqslant v
\end{array}\right. ; \\[4mm]
s>0,\, \tau<0   &  \Rightarrow & \left\{ \begin{array}{ll} \max\{-\varkappa,m \} \leqslant t_1 \leqslant \min\{\varkappa, n \}, & |t_1|\leqslant v \\
\max\{ -\varkappa,m,n  \} \leqslant t_2 \leqslant \varkappa , &
|t_2| \geqslant v
\end{array}\right. ; \\[4mm]
s<0, \, \tau>0, \,\sigma >0  &  \Rightarrow & \left\{
\begin{array}{ll} \max\{ m,n\} \leqslant t_1 \leqslant -\varkappa,
& |t_1|\leqslant v \\
m \leqslant t_2 \leqslant \min\{ -\varkappa, n \}, & |t_2| \geqslant
v
\end{array}\right. ;\\[4mm]
s>0,\, \tau>0,\, \sigma >0    &  \Rightarrow & \left\{
\begin{array}{ll}
 m \leqslant t_1 \leqslant \min\{-\varkappa, n \},
& |t_1|\leqslant v \\
\max\{ m,n \} \leqslant t_2 \leqslant -\varkappa, & |t_2| \geqslant
v
\end{array}\right. .
\end{array}
\end{equation}

\begin{figure}[htp]
\centering
\includegraphics[width=60mm,keepaspectratio]{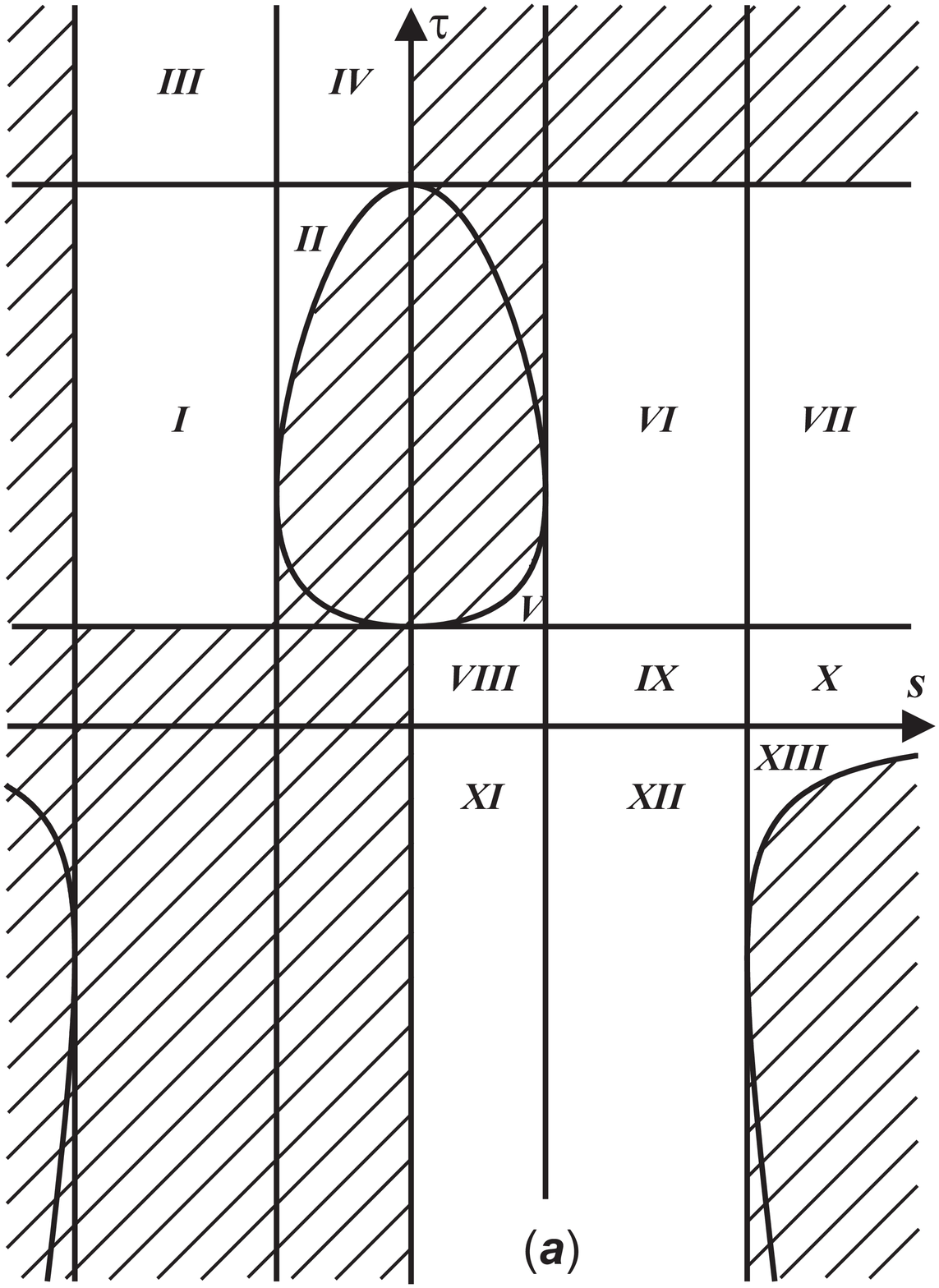}\hspace{10mm}
\includegraphics[width=60mm,keepaspectratio]{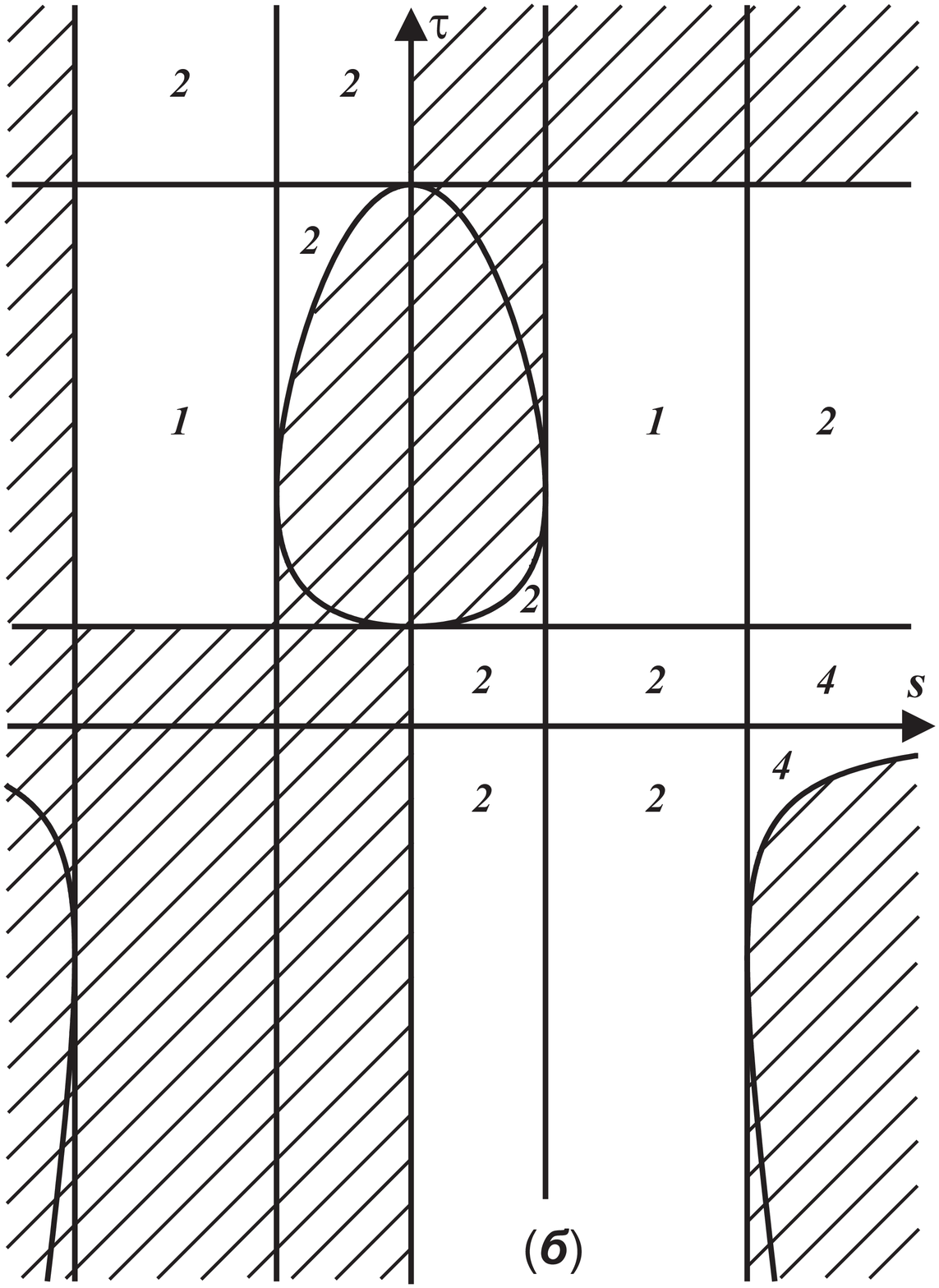}
\caption{Разделяющие кривые, кодировка областей и количество
торов.}\label{fig_harl1}
\end{figure}

При $\sigma<0$ всегда $\tau>0$. При этом $\varkappa$ чисто мнимое,
поэтому возникает условие (\ref{neq5_11}). Следовательно, аргументы
$u_1,u_2,u_6,u_7$ не влияют на результат. В импликациях
\eqref{neq5_14} существенны лишь две последних. С учетом возможностей
(\ref{neq5_13}), для редуцированного вектора ${\tilde{\mathbf{u}}}
=(u_3,u_4,u_5,u_8,u_9,u_{10})$ получаем решения
\begin{equation}\notag
\begin{array}{lll}
s \tau < 0 & \Rightarrow  & {\tilde{\mathbf{u}}} \in \{000011, 110011 \};\\
s \tau > 0 & \Rightarrow  & {\tilde{\mathbf{u}}} \in \{011000,
011110 \}.
\end{array}
\end{equation}
Отсюда для $s < 0$ имеем
\begin{equation}\label{neq5_16}
\left\{ \begin{array}{ll} t_1 \geqslant \max\{m,n\} , & |t_1|\leqslant v \\
t_2 \in [m,n], & |t_2| \geqslant v
\end{array}\right.
\end{equation}
или
\begin{equation}\label{neq5_17}
\left\{ \begin{array}{ll} t_1 \leqslant \min\{m,n\} , & |t_1|\leqslant v \\
t_2 \in [m,n], & |t_2| \geqslant v
\end{array}\right. .
\end{equation}
Для $s > 0$ варианты таковы
\begin{equation}\notag
\left\{ \begin{array}{ll} t_1 \in [m,n], & |t_1| \leqslant v \\
t_2 \geqslant \max\{m,n\} , & |t_2|\geqslant v
\end{array}\right., \qquad \left\{ \begin{array}{ll} t_1 \in [m,n], & |t_1| \leqslant v \\
t_2 \leqslant \min\{m,n\} , & |t_2|\geqslant v
\end{array}\right. .
\end{equation}
В этом случае они не являются взаимоисключающими, так как при
$s\tau>0$ переменная $t_2$ периодически <<пересекает>> бесконечно
удаленную точку (под радикалов в правой части уравнений~\eqref{neq5_5}~-- многочлены четной степени с положительным старшим
коэффициентом, и лишь переменная $t_2$, согласно договоренности,
может быть неограниченной). Поэтому их следует объединить в один
\begin{equation}\label{neq5_18}
\left\{ \begin{array}{l} t_1 \in [m,n], \qquad |t_1| \leqslant v \\
t_2 \in [-\infty, \min\{m,n,-v\}] \cup [\max \{m,n,v\},+\infty]
\end{array}\right. .
\end{equation}
Остается заметить, что при $\sigma <0$ имеются соотношения между
$m,n,v$, определяемые только значением параметра $s$. В
этом случае $\tau >0$ и $|n|<|m|$. Из определений
параметров (\ref{neq5_2}), (\ref{neq5_12}) получаем:
\begin{equation}\label{neq5_19}
\begin{array}{rcl}
|s|>a \Rightarrow & |n|<|m|<|v| & \Rightarrow -v <m<n<v;
\\
b<|s|<a \Rightarrow & |n|<|v|< |m| & \Rightarrow m<-v <n<v;
\\
|s|<b \Rightarrow & |v|<|n|<|m| & \Rightarrow m<-v <v<n.
\end{array}
\end{equation}
Вариант \eqref{neq5_17} несовместен ни с одним из последних.
Перекрестное сопоставление \eqref{neq5_16}, \eqref{neq5_18} с
возможностями \eqref{neq5_19} дает лишь пять допустимых вариантов при
$\sigma<0$ (по поводу обозначений см. замечание~\ref{rem81}):
\begin{equation}\label{neq5_20}
\begin{array}{lcll}
\sigma<0, \,\tau > 0,\, -a<s<-b  &  \Rightarrow &
t_1\in [\,n,v\,],& t_2 \in [\,m,-v\,];\\
\sigma<0,\, \tau > 0,\, -b<s<0  &  \Rightarrow & t_1\in [\,-v,v\,],& t_2 \in [\,m,n\,];\\
\sigma<0,\, \tau > 0, \,s> a &  \Rightarrow & t_1\in [\,m,n\,],& t_2 \in [\,v \,(\pm \infty) -v \,];\\
\sigma<0,\, \tau > 0, \,b<s<a  &  \Rightarrow & t_1\in [\,-v,n\,],& t_2 \in [\,v \,(\pm \infty)\, m \,];\\
\sigma<0,\, \tau > 0,\, 0<s<b  &  \Rightarrow & t_1\in [\,-v,v\,],& t_2 \in [\,n \,(\pm \infty)\, m \,].\\
\end{array}
\end{equation}

Подводя итоги, получим, что из 32 областей, на которые
дискриминантное множество максимального многочлена делит плоскость
$(s,\tau)$, согласно (\ref{neq5_15}), (\ref{neq5_20}) непустые
достижимые области для переменных $(t_1,t_2)$ имеются лишь в
тринадцати (см. рис.~\ref{fig_harl1},{\it а}). Вся информация
приведена в табл.~2. О последнем столбце речь пойдет ниже.

\begin{table}[htp]
\centering
\begin{tabular}{|c| c| c| c| c|c|c|}
\multicolumn{5}{r}{\fts{Таблица 2}}\\
\hline
\begin{tabular}{c}Номер\\области\end{tabular}&\begin{tabular}{c}Корни\end{tabular}&\begin{tabular}{c}Область\\изменения $t_1$\end{tabular}
&\begin{tabular}{c}Область\\изменения $t_2$\end{tabular}&
\begin{tabular}{c}Вторая\\группа\end{tabular}\\
\hline \ru  \ts{I} & $m<-v<n<v$ & $[m,-v]$ & $[n,v]$ & {$3\,5\,9\,{10}^*$} \\
\hline \ru \ts{II} & $m<n<-v<v$ & $[m,n]$ & $[-v,v]$ & {$3\,4\, {10}^*$}\\
\hline \ru \ts{III} & $m<-v<n<-\vk<\vk<v$ & $[m,-v]$ & $[n,-\vk]$ & {$3\,5\,6\,9$}\\
\hline \ru \ts{IV} & $m<n<-v<-\vk<\vk<v$ & $[m,n]$ & $[-v,-\vk]$ & {$3\,4\,6\,{10}$}\\
\hline \ru \ts{V} & $m<-v<v<n$ & $[-v,v]$ & $[n \,(\pm \infty)\, m ]$  &  {$5\,8\,9^*$} \\
\hline \ru \ts{VI} & $m<-v<n<v$ & $[-v,n]$ & $[v \,(\pm \infty)\,m]$  & {$4\,5\,8\,{10}^*$}\\
\hline \ru \ts{VII} & $-v<m<n<v$ & $[m,n]$ & $[v\,(\pm \infty)\,-v]$  & {$3\,4\,{10}^*$}\\
\hline \ru \ts{VIII} & $m<-v<-\vk<\vk<v<n$ & $[\vk,v]$ & $[n\,(\pm \infty)\,m]$ &   {$2\,5\,8\,{9}$}\\
\hline \ru \ts{IX} & $m<-v<-\vk<\vk<n<v$ & $[\vk,n]$ & $[v\,(\pm \infty)\,m]$ &   {$2\,4\,8\,{10}$}\\
\hline \ru \ts{X} & $-v<m<-\vk<\vk<n<v$ & $[\vk,n]$ & $[v\,(\pm \infty)\,-v]$ &  {$2\,4\,{10}$}\\
\hline \ru \ts{XI} & $-\vk<-v<m<n<v<\vk$ & $[m,n]$ & $[v,\vk]$ & {$3\,4\,7\,{10}$}\\
\hline \ru \ts{XII} & $-\vk<-v<m<v<n<\vk$ & $[m,v]$ & $[n,\vk]$ & {$3\,5\,7\,{9}$}\\
\hline \ru \ts{XIII} & $-\vk<m<-v<v<n<\vk$ & $[-v,v]$ & $[n,\vk]$ & {$5\,7\,9$}\\
\hline
\end{tabular}
\end{table}

\begin{table}[htp]
\centering
\begin{tabular}{|c| l|c|c|}
\multicolumn{4}{r}{\fts{Таблица 3}}\\
\hline
\begin{tabular}{c}Номер\\области\end{tabular} & \multicolumn{1}{c|}{Выражения компонент}&\begin{tabular}{c}$(P,Q)$\end{tabular}
&\begin{tabular}{c}$c(\mbf{f})$\end{tabular} \\
\hline \ru  $\ts{I}^*$ & $\quad u_1 \oplus u_2 \oplus u_6 \oplus u_7$& (0,0)&{1}\\
\hline \ru $\ts{II}^*$ & $\left\{ \begin{array}{l}u_1 \oplus u_2
\oplus u_6\oplus u_7\\u_5 \oplus u_8 \oplus u_9 \oplus
u_{11}\end{array}\right.$& (1,0)&{2}\\
\hline \ru $\ts{III}\;$ & $\quad u_1 \oplus u_4 \oplus u_7 \oplus u_8 \oplus u_{10}\oplus u_{11}$ & (1,0)&{2}\\
\hline \ru $\ts{IV}\;$ & $\quad u_1 \oplus u_2 \oplus u_5 \oplus u_8 \oplus u_{9}\oplus u_{11}$ & (1,0)&{2}\\
\hline \ru $\ts{V}^*$ & $\left\{ \begin{array}{l}u_1 \oplus u_2
\oplus
u_6 \oplus u_7\\
u_3 \oplus u_4 \oplus u_6 \oplus u_7 \oplus u_{10} \oplus u_{11} \end{array}\right.$&  {(1,0)}&{2}\\
\hline \ru $\ts{VI}^*$ & $\quad u_1 \oplus u_2 \oplus u_6 \oplus u_7 $ &  {(0,0)}&{1}\\
\hline \ru $\ts{VII}^*$ & $\left\{ \begin{array}{l}u_1 \oplus u_2
\oplus u_6\oplus u_7 \\u_5 \oplus u_6 \oplus u_7 \oplus u_8 \oplus u_{9} \oplus u_{11} \end{array}\right.$ &  {(1,0)}&{2}\\
\hline \ru $\ts{VIII}\;$ & $\quad u_3 \oplus u_4 \oplus u_6 \oplus u_7 $&  {(1,0)}&{2}\\
\hline \ru $\ts{IX}\;$ & $\quad u_1 \oplus u_3 \oplus u_5 \oplus u_7 \oplus u_{9}\oplus u_{11}$&  {(1,0)}&{2}\\

\hline \ru $\ts{X}$ & $\left\{ \begin{array}{l} u_1 \oplus u_3
\oplus u_6 \oplus u_8\\u_5 \oplus u_6 \oplus u_7 \oplus u_8 \oplus u_{9} \oplus u_{11} \end{array}\right.$ &  {(2,0)} & {4} \\

\hline \ru $\ts{XI}$ & $\quad u_1 \oplus u_2 \oplus u_5 \oplus u_8 \oplus u_{9}\oplus u_{11}$& (1,0)&{2}\\
\hline \ru $\ts{XII}$ & $\quad u_2 \oplus u_4 \oplus u_6 \oplus u_8 \oplus u_{10}\oplus u_{11}$ & (1,0)&{2} \\
\hline \ru $\ts{XIII}$ & $\left\{ \begin{array}{l}u_1 \oplus u_3\oplus u_6\oplus u_8\\u_2 \oplus u_4 \oplus u_6 \oplus u_8 \oplus u_{10} \oplus u_{11} \end{array}\right.$& (2,0)&{4}\\
\hline
\end{tabular}
\end{table}
Отметим, что при решении задачи определения достижимых и допустимых
областей к формулам надстройки (\ref{neq5_4}) мы обратились
единственный раз при выписывании компонент линейной булевой
вектор-функции. После этого, пользуясь лишь формальным аппаратом
редукции двоичных матриц и свойствами параметров, мы доказали
следующие утверждения.
\begin{theorem}\label{th91}
Обозначим
$$
\theta(\tau)=\ds{\frac{1}{2}\sqrt{2p^2-\frac{\tau^2+r^4}{\tau}}}.
$$
Образ отображения $(\ref{neq5_1})$ в плоскости $(s,\tau)$
определяется неравенствами:
$$
\begin{array}{ll}
0<s \leqslant a, & \tau \leqslant -r^2;\\
0<s \leqslant \theta(\tau), & -r^2 \leqslant \tau <0 ;\\
0<s < + \infty, & 0 \leqslant \tau < (a-b)^2 ;\\
(-a \leqslant s \leqslant -b) \cup (\theta(\tau)\leqslant s <+\infty) , & (a-b)^2 \leqslant \tau \leqslant r^2 ;\\
(-a \leqslant s \leqslant -\theta(\tau)) \cup (b \leqslant s <+\infty) , & r^2  \leqslant \tau \leqslant (a+b)^2 ;\\
-a \leqslant s <0  , & \tau  > (a+b)^2.
\end{array}
$$
\end{theorem}

\begin{theorem}\label{th92}
Бифуркационная диаграмма отображения $(\ref{neq5_1})$ состоит из
следующих множеств:
$$
\begin{array}{ll}
\tau=0, & s \in (0,+\infty);\\
\tau=(a-b)^2, & s \in [-a,-b] \cup (0,+\infty); \\
\tau=(a+b)^2, & s \in [-a,0] \cup (b,+\infty); \\
s=-a, & \tau \in [(a-b)^2,+\infty); \\
s=-b, & \tau \in [(a-b)^2,+\infty); \\
s=-\theta(\tau), & \tau \in (r^2,(a+b)^2); \\
s=\theta(\tau), & \tau \in (-r^2,0) \cup ((a-b)^2,r^2); \\
s=b, & \tau \in (-\infty,(a+b)^2]; \\
s=a, & \tau \in (-\infty,(a+b)^2].
\end{array}
$$
\end{theorem}

Обратимся к задаче исследования топологии интегральных многообразий.
Для того чтобы установить количество торов Лиувилля в составе
$\mF_{\mbf{f}}$, воспользуемся редуцированной функцией
(\ref{neq5_9}). В последнем столбце табл.~2 приведена
информация о радикалах второй группы (в данной задаче они составляют
меньшинство). Звездочка означает необходимость фильтрации наборов
аргументов в соответствии с условием (\ref{neq5_11}).

В областях \ts{V}--\ts{X} переменная $t_2$ пересекает бесконечность.
Здесь удобно поступить подобно тому, как это делалось в
\cite{KhNDnew} для классического случая Ковалевской. Разница в том,
что среди радикалов $R_{2\alpha}$ здесь нельзя указать такого,
который всегда бы относился к первой группе. Кроме того,
максимальный многочлен имеет четную степень, поэтому $\infty$ не
является точкой ветвления для $\sqrt{V}$. Поступим так. Поскольку
всегда $|t_2| \geqslant v$, то $t_2 \ne 0$. Положим
$$
R_{2 \alpha}^*= R_{2 \alpha}/ \sqrt{t_2}, \qquad u_{5+\alpha}=\lsgn
R_{2 \alpha}^* \qquad (\alpha=1,\ldots, 5).
$$
После перехода к новым
радикалам особенность $t_2=\infty$ в выражениях (\ref{neq5_4}) исчезает, а при
пересечении этого значения теперь никакие подкоренные выражения в
ноль не обращаются. Поэтому аргументы $u_\alpha$ для $\alpha
>5$ необходимо относить к соответствующей группе, рассматривая лишь конечные границы промежутка изменения $t_2$.

Теперь в соответствии со списком аргументов второй группы выполняем
эквивалентные преобразования с основной матрицей $C$ в формуле
(\ref{neq5_8}). А именно, если аргумент $u_\alpha$ относится ко
второй группе, то с помощью строки, где он присутствует, обнуляем
все остальные элементы этого столбца и по теореме~\ref{theo2}
исключаем аргумент и единственную содержащую его компоненту функции.
Остающиеся компоненты показаны в табл.~3. Особенность матрицы
такова, что во всех случаях \ts{I}--\ts{XIII} все аргументы второй
группы позволяют исключить ровно одну компоненту. Таким образом,
ранг оставшейся матрицы равен <<пять минус количество аргументов
второй группы>>, причем это всегда есть первое число в паре рангов
$(P,Q)$ (см. теорему~\ref{theo3}), так как аргументов второй группы
не остается. Соответственно, всегда в такой паре $Q=0$
(предпоследний столбец в табл.~3). В случаях со звездочкой одна из
оставшихся компонент всегда имеет вид
$$
u_1 \oplus u_2 \oplus u_6 \oplus u_7,
$$
то есть является константой в силу (\ref{neq5_11}), поэтому ранг надо
уменьшить на единицу. Через $c(\mbf{f})$ здесь и далее
обозначим количество классов эквивалентности в пространстве аргументов первой группы для заданного значения
$\mbf{f}=(s,\tau)$ постоянных первых интегралов. В общем случае $c(\mbf{f})$ зависит еще от выбора связной области осцилляции переменных, но в рассматриваемой задаче при любых допустимых $\mbf{f}$ такая область одна (см. табл.~2 с учетом прохождения значения $t_2=\infty$ в конечные моменты времени). Поэтому $c(\mbf{f})$  совпадает с количеством
связных компонент интегральной поверхности $\mF_\mbf{f}$. Окончательно, число торов Лиувилля равно $2^P$ и указано
в последнем столбце табл.~3. Результат по количеству торов показан и на рис.~\ref{fig_harl1},{\it б}. Исследование
регулярной фазовой топологии задачи завершено.

Рассмотрим критические случаи --- отрезки бифуркационного множества
между областями. Поскольку наша цель --- продемонстрировать технику,
то для краткости известные переходы рассматривать не будем, а
применим результаты предыдущих разделов. Действительно, случаи
$\chi=0$ отвечают в точности особым периодическим решениям случая
Богоявленского, причем константа интеграла $s$ --- это использованный
в разделе~\ref{sec3} параметр $s$. Кривая (\ref{neq5_6}) накрывается
кривыми \eqref{neq3_1} дважды ($\pm f$). Поэтому на ее участках имеем
следующие критические поверхности: $2 S^1$ при $s\in (-b,0) \cup
(0,b)$, $4 S^1$ при $s\in (a,+\infty)$. Полупрямая $\tau=0, s>0$
отвечает пересечению со второй критической системой, в которой этому
множеству поверхностей соответствует подробно исследованный случай
$\ell=0, m<0$ и при этом связь параметров такова $m=-1/(2s)$.
Поэтому в рассматриваемой системе интегральные поверхности для
$\tau=0$ таковы: $2\bT^2$ при $s\in (0,b) \cup (b,a)$, $4\bT^2$ при
$s\in (a,+\infty)$. Этот случай является <<полурегулярным>>, так как
реальных бифуркаций не происходит, возникает лишь особенность в
выражениях надстройки, которую можно устранить другой заменой
переменных.

Для примера рассмотрим более подробно переходы, возникающие при
круговом обходе критической точки типа <<седло-седло>>.

\begin{figure}[htp]
\centering
\includegraphics[width=60mm, keepaspectratio]{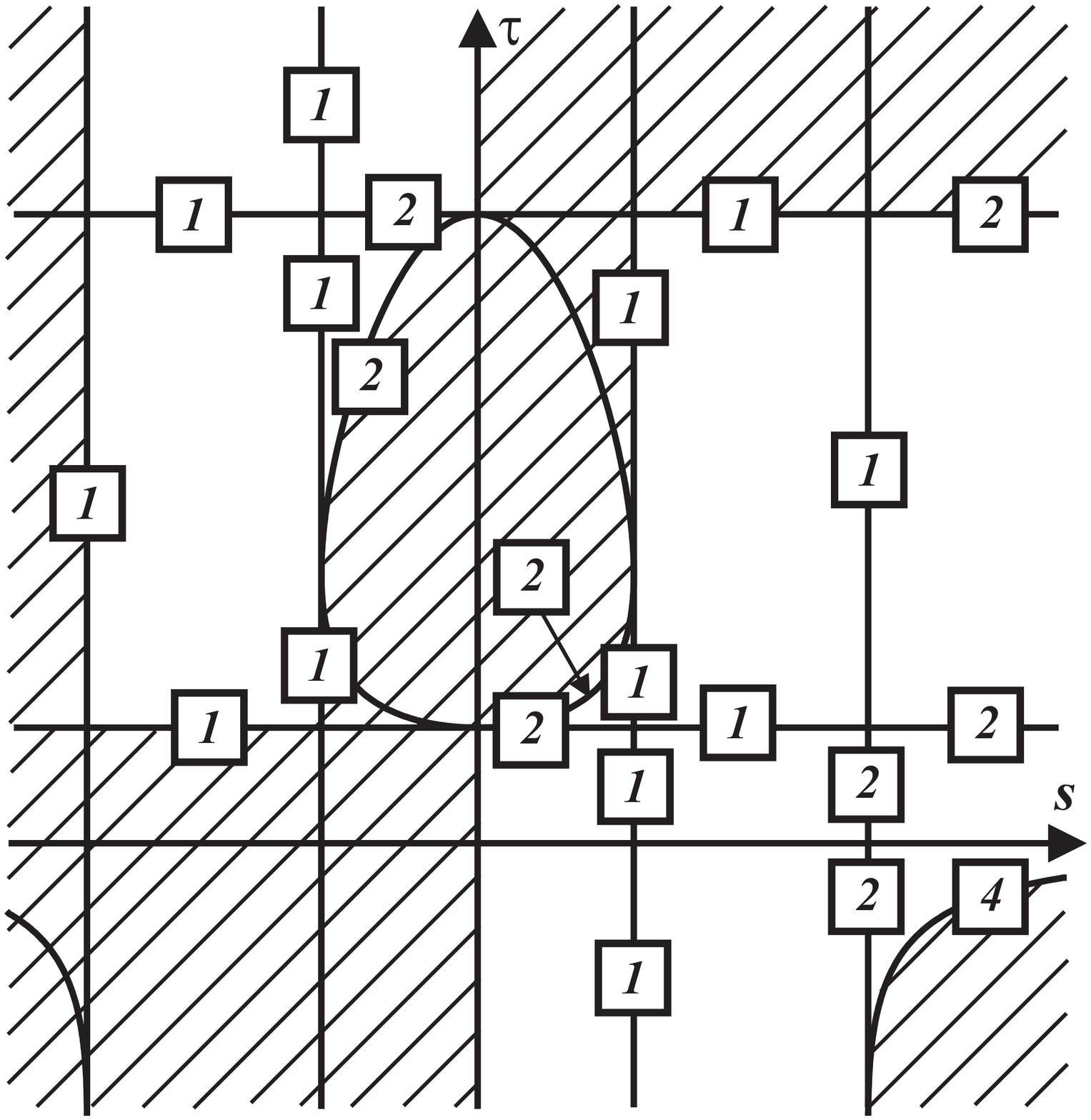}
\caption{Связность критических поверхностей.}\label{fig_harl2}
\end{figure}

При переходе $\ts{I}{\to}\ts{II}$ возникает кратный корень $-v=n$.
Во вторую группу попадают радикалы $3\,4\,5\,9\,10^*$. Переменная
$u_5$ позволяет исключить единственную нетривиальную компоненту
области $\ts{II}^*$ (см. табл.~3). Получаем пару $(P,Q)=(0,0)$,
поэтому $c(\mbf{f})=1$. Но перестройка здесь $\bT^2 \to 2\bT^2$.
Следовательно, $\mF_{s,\tau}=(S^1 \vee S^1)\times S^1$.

При переходе $\ts{I}{\to}\ts{III}$ возникает кратный корень
$\varkappa=0$. Во вторую группу попадают радикалы $3\,5\,6\,7\,9$.
Переменная $u_7$ позволяет исключить единственную нетривиальную
компоненту области $\ts{III}$. Получаем $(P,Q)=(0,0)$,
$c(\mbf{f})=1$. Перестройка $\bT^2 \to 2\bT^2$, значит,
$\mF_{s,\tau}=(S^1 \vee S^1)\times S^1$.

При переходе $\ts{III}{\to}\ts{IV}$ возникает кратный корень $-v=n$
строго между $m$ и $-\varkappa$. Во вторую группу попадают радикалы
$3\,4\,5\,6\,8\,10$. По сравнению с регулярной областью \ts{III} добавились
аргументы $u_4,u_8$, любой из них позволяет исключить единственную
компоненту функции. По сравнению с регулярной областью \ts{IV} добавились
аргументы $u_5,u_8$, и снова любой из них позволяет исключить
единственную компоненту функции. В результате $(P,Q)=(0,0)$,
$c(\mbf{f})=1$. Перестройка $2\bT^2 \to 2\bT^2$, значит,
$\mF_{s,\tau}=(S^1 \ddot \vee S^1)\times S^1$, где $S^1 \ddot \vee S^1$ --- пара окружностей, пересекающихся по двум точкам.

При переходе $\ts{II}{\to}\ts{IV}$ возникает кратный корень
$\varkappa=0$ между $-v$ и $v$. Во вторую группу попадают радикалы
$3\,4\,6\,7\,10$. По сравнению с регулярной областью \ts{II} добавились
аргументы $u_6,u_7$, которые входили в компоненту-константу, поэтому
это не повлияло на результат. По сравнению с регулярной областью \ts{IV}
добавился аргумент $u_7$, и он не входит в оставшуюся согласно
табл.~3 компоненту функции. Поэтому результат такой же, как и в
самих областях --- $(P,Q)=(1,0)$, $c(\mbf{f})=2$. Однако, здесь
перестройка по количеству торов такая же, как в предыдущем случае
$2\bT^2 \to 2\bT^2$. Поэтому критическая поверхность из двух
компонент есть $\mF_{s,\tau}=2(S^1 \vee S^1)*S^1$ (каждая компонента --- нетривиальное
расслоение над <<восьмеркой>> со слоем окружность).

Аналогично рассматриваются и все остальные случаи. На
рис.~\ref{fig_harl2} показано количество связных компонент
критических поверхностей для всех участков бифуркационной диаграммы,
из чего, благодаря найденному ранее количеству торов Лиувилля в
регулярных областях однозначно восстанавливается и топологический
тип любой критической поверхности.

{\bf Благодарности.} Работа выполнена при финансовой поддержке РФФИ
(проект № 10-01-00043).

\end{document}